\documentclass[12pt]{iopart}
\usepackage[utf8]{inputenc}
\usepackage[american,british]{babel}
\usepackage[T1]{fontenc}
\usepackage[pdftex]{graphicx}  
\usepackage{graphicx, xcolor}
\usepackage{dcolumn}
\usepackage{braket}
\usepackage{bm}
\expandafter\let\csname equation*\endcsname\relax
\expandafter\let\csname endequation*\endcsname\relax
\DeclareUnicodeCharacter{0301}{\'{e}}

\usepackage{amsmath,amsthm,amssymb}
\usepackage{mathtools}
\usepackage{color}
\usepackage{verbatim}
\usepackage{comment}
\usepackage{lipsum}
\usepackage[colorlinks,linkcolor=blue,citecolor=blue,urlcolor=blue]{hyperref}
\usepackage{mathrsfs}
\usepackage{subcaption}

\usepackage[backend=biber, style=numeric-comp, doi=true, isbn=false, url=false, eprint=false, sorting=none,maxnames=10,firstinits=true]{biblatex}
\usepackage{physics}
\usepackage{mathtools}
\newcommand{\mean}[1]{\left\langle #1 \right\rangle}
\newcommand{\be}{\begin{equation}}
\newcommand{\ee}{\end{equation}}
\newcommand{\bea}{\begin{eqnarray}}
\newcommand{\eea}{\end{eqnarray}}

\newcommand{\Pnoclick}{P^{(0)}}
\newcommand{\Pclick}{P^{(n>0)}}
\newcommand{\moment}[3][x]{\mean{#1^{#2}}_{P_{#3}}}

\begin{document}

\title{%Investigating the Full Counting Statistics Beyond Lindblad Dynamics in Continuously Monitored Quantum Systems
Continuously Monitored Quantum Systems Beyond Lindblad Dynamics}

\author{Guglielmo Lami$^{1,*}$, Alessandro Santini$^{1,\dagger}$, Mario Collura$^{1,2,\ddagger}$}

\address{$^1$ SISSA, via Bonomea 265, 34136 Trieste, Italy}
\address{$^2$ INFN, via Bonomea 265, 34136 Trieste, Italy}

\ead{$^*$glami@sissa.it}
\ead{$^\dagger$asantini@sissa.it}
\ead{$^\ddagger$mcollura@sissa.it}

\vspace{10pt}
\begin{indented}
\item[]\today
\end{indented}

\begin{abstract}
    The dynamics of a quantum system, undergoing unitary evolution and continuous monitoring, can be described in term of quantum trajectories. Although the averaged state fully characterises expectation values, the entire ensamble of stochastic trajectories goes beyond simple linear observables, keeping a more attentive description of the entire dynamics.  Here we go beyond the Lindblad dynamics and study the probability distribution of the expectation value of a given observable over the possible quantum trajectories. The measurements are applied to the entire system, having the effect of projecting the system into a product state.   
    We develop an analytical tool to evaluate this probability distribution at any time $t$. 
    %The method relies on splitting the contributions due to to the trajectories with fixed number of measurements $n$ and rewriting each contribution in term of an effective transition matrix. 
    We illustrate our approach by analyzing two paradigmatic examples: a single qubit subjected to magnetization measurements, and a free hopping particle subjected to position measurements. 
    %Our analysis provides new insights of the behavior of continuously monitored quantum systems and gives access to information that are not encoded in the lindbladian description of the time-evolution.  
\end{abstract}

%%%%%%%%%%%%%%%%%%%
\section{Introduction}
Quantum mechanics is a fundamental milestone of the human comprehension of natural world \cite{penrose2004road}. One of its most enigmatic and controversial features is the role played by measurements \cite{penrose2004road,Neumann2018,Zurek1983Book,holevo2003statistical,Bassi2013RevModPhys}. While in a classical perspective measurements are trivially a way of extracting information from a system, in quantum mechanics the meaning of measurements is much more profound \cite{wiseman_milburn_2009}. When a quantum system is measured, its wave function undergoes a non-deterministic collapse and the system is projected into a specific state \cite{Bohr1928}. Recent technological advancements have enabled increasingly accurate and fast measurements on quantum systems \cite{Katz2006Science,Ibarcq2016PRX}. These advances have opened up new avenues for exploring the fundamental principles of quantum mechanics \cite{Greiner2015Science, smith2019simulating, Santini2022PRA} and have led to the development of novel applications in fields such as quantum information processing \cite{Preskill2018quantumcomputingin,DelCampo2019PRL} and quantum thermodynamics \cite{Campisi2010PRL,Solfanelli_2019,Buffoni2019PRL,gherardini1,gherardini2,Solfanelli2021PRXQ,gherardini3,Manzano2022AVSQuantum,Santini2022Arxiv}. However, the dynamics of continuously monitored quantum systems are often difficult to analyze, and there is a need for theoretical approaches that can provide insights into the behavior of these systems

Great interest has been shown in measurement-induced criticality, which has emerged as a prominent topic in the study of continuously monitored quantum systems \cite{DeLuca2019,szyniszewski2019entanglement,szyniszewski2020universality,turkeshi2020measurement,Vasseur2020PRB,alberton2020trajectory,turkeshi2021measurement,turkeshi2021measurement2,botzung2021engineered,buchhold2021effective,muller2021measurement,biella2021many,Sierant2022dissipativefloquet,turkeshi2022entanglement,sharma2022measurement,sierant2022universal,coppola2022growth,piccitto2022entanglement,Turkeshi2022PRB,Boorman2022PRB,tirrito2023counting,piccitto2023entanglement,Romito2023Scipost,paviglianiti2023multipartite,zerba2023measurement,Minoguchi_2022}. Indeed, continuous measurements of a quantum system can create a feedback loop that leads to a non-equilibrium steady state. In this steady state, the system can exhibit different phases depending on the strength and type of measurement. For example, in the quantum Zeno phase, frequent measurements can suppress quantum fluctuations, leading to a phase where the system behaves as if it were frozen \cite{Muller2017ADP,Romito2020PRR}. Conversely, when measurements are less frequent or weaker, the system can enter a volume law phase, where the entanglement entropy grows linearly with the system size. Nonetheless, in experiments on measurement-induced criticality, post-selection is often required to reveal the underlying physics hidden in quantum trajectories \cite{ippoliti2021postselection,koh2022experimental_PostSelection}.

In this article, we consider quantum systems that are coupled to a measuring apparatus, examining their evolution according to their unitary dynamics, which is interrupted by projective measurements with a fixed rate $\gamma$. The measurements are applied to the entire Hilbert space, thus projecting the system onto an %product state 
eigenstate of the measured observable. To illustrate our approach, we derive analytical results for two paradigmatic examples: a single qubit measuring its magnetization and a free hopping particle measuring its position. We therefore provide an exact method for computing the probability distribution of the expectation value of observables averaged over the set of quantum trajectories.

The approach we present here goes beyond a Lindblad master equation description. Indeed, the Lindblad equation describes the dynamics of a quantum system subjected to quantum jumps by considering the density matrix averaged over all possible outcomes, which is a non-selective state \cite{wiseman1996quantum,plenio1998quantum,breuer2002theory,wiseman_milburn_2009}. In the case of a continuously measured system, the Lindblad approach neglects the outcome of the measurements and averages over all possible outcomes. In contrast, our approach is selective, as we keep the information contained in individual quantum trajectories. This allows us to uncover the physics that is neglected in the average state.
%By utilizing this powerful tool, we can gain a deeper comprehension of the physics concealed within quantum trajectories. {\color{red} Ale: forse ci stanno due parole sul Beyond Lindblad?}

\section{Protocol}\label{sec:protocol}
\begin{figure}
    \centering
\includegraphics[width=.75\linewidth]{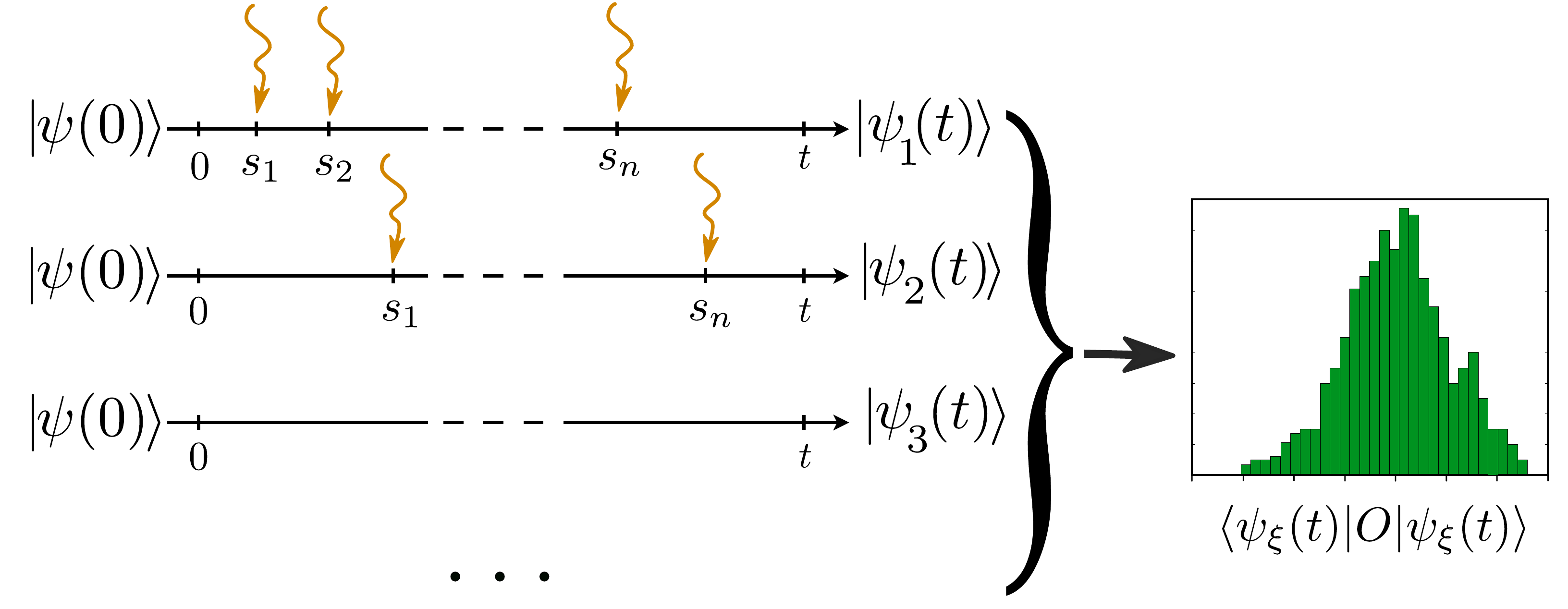}
    \caption{Scheme of the dynamical protocol. We consider a system prepared in a fixed initial state $\ket{\psi(0)} = \ket{a_0}$, undergoing random projective measurements of an observable $A$ (see Eq.~\eqref{eq:Aop}). Measurements occur with a fixed rate $\gamma$ at times $s_1 < s_2 <  ... $, whereas in the intervals $(s_j, s_{j+1})$ the system follows an unitary time evolution. Different quantum trajectories are labelled by the integer $\xi$. The ensemble of states $\ket{\psi_{\xi}(t)}$ defines the probability distribution function of $\expval{ O}{\psi_\xi(t)}$ (right).}
    \label{fig:my_label}
\end{figure}

Let us consider an $N$-level quantum system described by a time-independent hamiltonian $H$,
%\begin{equation}
%     H = \sum_{k=1}^N E_{k} \dyad{E_k},
%\end{equation}
whose unitary evolution is governed by $U(t)=e^{-i H t}$. In addition, all along the evolution, we couple the system to a measuring apparatus that project, with a fixed measurement rate $\gamma$, the evolved state in to an eigenstate of the observable
\begin{equation}\label{eq:Aop}
     A = \sum_{a=1}^N \nu_{a}\dyad{a},
\end{equation}
such that $[A,H]\neq 0$. At time $t=0$ the system has been prepared in a
fixed state $\ket{\psi(0)}$ corresponding to an eigenstate $\ket{a_0}$ of $A$. We are interested in evaluating the probability distribution of the expectation value of a generic %{\it diagonal} 
observable $O = \sum_a o_a \dyad{a}$ commuting with $A$, i.e.\ 
\begin{equation}
    P_O(x;t)= \overline{\delta\left(\expval{ O}{\psi_\xi(t)}-x\right)} \, , 
\end{equation}
with $t>0$. Here, the over-line is indicating the average over the quantum trajectories, which are labelled by the integer $\xi$. A sketch of our dynamical protocol is shown in Figure~\ref{fig:my_label}. We can split this average by putting apart the contributions for different $n$ clicks of the measuring apparatus. For each $n$ the system is projected in one of the eigenstates of $ A$ at times $\{s_j\}$, with $j=1,\dots,n$ such that $0 < s_1<\dots<s_n<t$. We define
\begin{equation}\label{eq:probability_ditribution}
    P_O(x;t)=\Pclick_O(x;t) + \Pnoclick_O(x;t) \, ,
\end{equation}
where $\Pnoclick$, $\Pclick$ are respectively the no-click ($n=0$) and click ($n>0$) contributions to the probability distribution of $\expval{ O}{\psi_\xi(t)}$. 
%Notice that we split apart the contribution in which the measurement apparatus never measure the system $\Pnoclick_O$ and the one in which $n>0$ projective measures are applied on the system $\Pclick_O$. 
Now we can write down these object as follows 
\begin{equation}
 \begin{cases}
\Pnoclick_O(x;t) =  
e^{-\gamma t}\mathscr{D} (x,a_0,t) \\
     \quad \\
    \Pclick_O(x;t) = \sum_{n=1}^\infty \gamma^n e^{-\gamma t} \sum_{a_1,\dots, a_n} \mathcal{T} \int_0^{t}\dd s_1\dots \dd s_n \, \,  p(x,t| s_n,a_n; \dots; s_1,a_1;0,a_0),
\end{cases}   
\end{equation}
where 
\begin{equation}
\mathscr{D} (x,a,t)\equiv\delta \left(x-\mel{a}{{U}^\dagger(t) {O}{U}(t)}{a} \right)
\end{equation}
and $p(x,t| s_n,a_n; ...; s_1,a_1;0,a_0)$ is the conditional probability density of getting the average value $x$ at the final time $t$, given that the system was found in the eigenstates $\{\ket{a_j}\}$ of ${A}$ at times $\{s_j\}$. $\mathcal{T}$ denotes the time-ordered product. 
The term $\gamma^n e^{-\gamma t}$
is the Poisson weight associated to the $n-$measurements events (the factor $1/n!$ is removed since inside the time-ordered integral the events labelled by $1,2... \, n$ occurs at fixed times $s_1 < s_2 <  ... \, < s_n$). Since the measurement process is Markovian, we have that
\begin{equation}
    p(x,t| s_n,a_n; \dots; s_1,a_1;0,a_0) =
     \mathscr{D}(x,a_n,t-s_n)
     p(s_n,a_n|s_{n-1},a_{n-1})
     \dots
     p(s_1,a_1|0,a_0), 
\end{equation}
where the delta contribution can be rewritten as
\be
\mathscr{D} (x,a,t) 
=\delta \big(x -
\sum_{a'=1}^{N} o_{a'} \, p(t,a'|0,a) 
\big),
\ee 
and gives the contribution to the probability due to the last time-interval $(s_n, t)$. In addition, the transition probabilities between eigenstates of ${A}$ read
\begin{equation}
   p(s',a'|s,a) \equiv \mathbb{T}_{a',a}(s'-s) =\abs{\mel{a'}{U(s'-s)}{a}}^2.
\end{equation}
Let us assume that the transition  matrix $\mathbb{T}$ can be diagonalised, i.e.\ $\mathbb{T}(s'-s) = \mathbb{V} \, \mathbb{D}(s'-s) \,\mathbb{V}^\dag$ with eigenvalues $\mathbb{D}_{\alpha,\beta}(t) = d_\alpha(t)\delta_{\alpha,\beta}$, and time-independent unitary matrix $\mathbb{V}$ whose columns are the orthonormal eigenvectors. Notice that, since $\mathbb{T}$ is a unitary-stochastic matrix, $d_1=1$ is the largest eigenvalue, meanwhile as a consequence of the Perron–Frobenius theorem $\abs{d_\alpha} \leq 1$ for $\alpha=2,\dots,N$, where the equality holds in case of degeneracy \cite{KarolZyczkowski_2003,gardiner2009stochastic}.
In addition, since $\sum_a \mathbb{T}_{a,b} = \sum_b \mathbb{T}_{a,b} = 1$, the eigenvector associated to largest eigenvalue is simply given by $\mathbb{V}_{a,1} = 1/\sqrt{N}$.
Now let us focus on the nontrivial  $\Pclick_O(x;t)$. For each order $n$ and fixed states $\{a_0, a_n\}$, the sum over all the the possible intermediate outcomes can be rewritten as
\begin{equation}
     \sum_{a_1,\dots,a_{n-1}} 
     \mathbb{T}_{a_{n},a_{n-1}}(s_n-s_{n-1}) 
      \cdots
      \mathbb{T}_{a_1,a_0}(s_1) = \sum_{\alpha=1}^{N}\mathbb{V}_{a_n,\alpha} 
     \prod_{j=1}^{n} d_{\alpha}(s_{j}-s_{j-1})
     (\mathbb{V}^{\dag})_{\alpha,a_0},
\end{equation}
with starting time $s_0=0$. We thus get
\begin{equation}
     \Pclick_O(x;t) = e^{-\gamma t}\sum_{n=1}^\infty  \left[ \int_0^t \gamma \dd s_n \sum_{a_n=1}^{N}
     \mathscr{D} (x,a_n,t-s_n) \sum_{\alpha=1}^N \mathbb{V}_{a_n,\alpha}
     I_{n,\alpha}(s_n)
     (\mathbb{V}^{\dag})_{\alpha,a_0} \right]\label{eq:generic_click}
\end{equation}
where we defined the time-ordered integrals
\begin{equation}
    I_{n,\alpha}(s_n) = \int_0^{s_n} \gamma \dd s_{n-1} \, d_\alpha(s_n-s_{n-1}) \, \dots \, \int_0^{s_2}\gamma \dd s_1 \, d_\alpha(s_2-s_1)d_\alpha(s_1),
\end{equation}
which satisfy the following recursive relations
\begin{equation}
\begin{dcases}
I_{1,\alpha}(s_1) = d_\alpha(s_1),\\
I_{n,\alpha}(s_n) = \int_0^{s_n}\gamma \dd s_{n-1} \, d_\alpha(s_n-s_{n-1}) I_{n-1,\alpha}(s_{n-1}), \quad n>1 .
\end{dcases}\label{eq:generic_recursion_relation}
\end{equation}
In Eq.~(\ref{eq:generic_click}) the sum over $n$ can be safely taken, since $s_n$ and $a_n$ are dummies variables. Thus, we can introduce
$I_{\alpha}(s) = \sum_{n=1}^{\infty} I_{n,\alpha}(s)$ which satisfy the following linear Volterra integral equation of the second kind
\begin{equation}\label{eq:volterra}
I_\alpha(s) - d_\alpha(s) = \int_0^s \gamma \dd s' \, d_\alpha(s-s')I_\alpha(s').
\end{equation}
Exploiting the Laplace transform of the integral kernel, i.e. 
$\mathscr{L}[d_\alpha](z) = \int_0^\infty e^{-zt} d_\alpha(t) \dd t$, we can easily solve the previous equation obtaining
\begin{equation}
    I_\alpha(s) = 
    \mathscr{L}^{-1}\left[
    \frac{\mathscr{L}[d_\alpha](z)}{1-\gamma\mathscr{L}[d_\alpha](z)}\right](s),
\end{equation}
where $\mathscr{L}^{-1}[\dots]$ is the inverse Laplace transform. As expected, it easy to show that
\begin{equation}
\mathscr{L}[I_\alpha](z) = 
 \frac{\mathscr{L}[d_\alpha](z)}{1-\gamma\mathscr{L}[d_\alpha](z)}
= \sum_{n=1}^{\infty} \gamma^{n-1}\mathscr{L}[d_\alpha]^n(z)
=\sum_{n=1}^{\infty}\mathscr{L}[I_{n,\alpha}](z),
\end{equation}
thus identifying 
$\mathscr{L}[I_{n,\alpha}](z) = \gamma^{n-1}\mathscr{L}[d_\alpha]^n(z)$.
%It is easy to evaluate $I_1(t)=e^{\gamma t}$.
We can finally collect all those results, further simplify and finding the following general expression for the probability distribution
\begin{equation}\label{eq:generic_click_2}
     \Pclick_O(x;t) = e^{-\gamma t} 
     \sum_{a=1}^{N} \sum_{\alpha=1}^N
     \int_0^t \gamma \dd s \,
     \mathscr{D} (x,a,t-s)  \mathbb{V}_{a,\alpha}
     I_{\alpha}(s)(\mathbb{V}^{\dag})_{\alpha,a_0}.
\end{equation}
Let us finally mention that using this expression we can easily compute the $k$-th moment of the distribution, defined as $\moment{k}{O} = \int \dd x \, x^k P_O (x;t) = \overline{\expval{O}{\psi_\xi(t)}^k}$. The first moment ($k=1$) is particularly simple, since it does reduce to the expectation value of $O$ over the averaged density matrix $\overline{\ket{\psi_{\xi}(t)}\bra{\psi_{\xi}(t)}}$ whose dynamics is fully described by the Lindblad equation (see \ref{app:moment_qm}). In particular, using that
$\sum_{b}\mathbb{T}_{a,b}(t-s) \mathbb{V}_{b,\alpha} = d_{\alpha}(t-s) \mathbb{V}_{a,\alpha}$,
we easily get
\be
\moment{}{O} = e^{-\gamma t} 
\sum_{a=1}^{N} o_{a} \, \mathbb{T}_{a,a_0}(t) +
e^{-\gamma t} 
\sum_{a=1}^{N} o_a 
\sum_{\alpha=1}^{N}  \mathbb{V}_{a,\alpha}
\left[\int_0^t \gamma \dd s \,
     d_\alpha(t-s) I_{\alpha}(s) \right]
(\mathbb{V}^{\dag})_{\alpha,a_0},
\ee
which can be further simplified exploiting Eq.~(\ref{eq:volterra}), finally obtaining
\be\label{eq:moment1}
\moment{}{O} = 
e^{-\gamma t} 
\sum_{a=1}^{N} o_a 
\sum_{\alpha=1}^{N}  \mathbb{V}_{a,\alpha}
I_{\alpha}(t)
(\mathbb{V}^{\dag})_{\alpha,a_0}.
\ee
Let us stress that the simplified expression in Eq.~\eqref{eq:moment1} applies only for averages of diagonal observables. Every time we are interested in higher moments or probability distribution of non-diagonal operators, the computation have to be carried out from scratch, basically starting from Eq.~\eqref{eq:generic_click_2}.
Finally, due to the properties of the transfer matrix $\mathbb{T}$, the stationary limit $t\to\infty$ of the previous average can be easily taken; in fact, only $\alpha=1$, with $I_{1}(t) = e^{\gamma t}$, will contribute to the sum over $\alpha$, 
leading to $\moment{}{O} \to \frac{1}{N} \sum_{a=1}^{N} o_a$,
where we used the fact that $\mathbb{V}_{a,1} = 1/\sqrt{N}$. As expected from the Lindblad dynamics, it does correspond to the expectation value of the operator $O$ over the infinite temperature state $\frac{1}{N} \sum_{a=1}^{N} \ket{a}\bra{a}$.

\section{Two-level system}\label{sec:single_spin}
We start by considering a continuously monitored two-level system. Despite their simplicity, two-level systems are the fundamental building block of the most studied quantum many-body systems and therefore understanding their behavior is crucial. The system consists of a single spin-$1/2$, whose 
unitary evolution is governed by the following hamiltonian\begin{equation}
    {H} = -J \sigma^x ,
\end{equation}
where $\sigma^\alpha$ are the Pauli matrices with $\alpha=x,y,z$ and $\comm{\sigma^\alpha}{\sigma^\beta} = 2i\epsilon_{\alpha\beta\gamma}\sigma^\gamma$. The system is continuously monitored along $z$ with a rate $\gamma$ (therefore ${A} = \sigma^z$). We will denote with $\ket{\sigma}=\ket{+1},\ket{-1}$ the eigenstates of $\sigma^z$ with eigenvalue $\sigma=\pm 1$ respectively. We consider a two-level spin which starts from $\ket{+1}$ and we want to evaluate
\begin{equation}
    P_{\sigma^z}(x;t)= \overline{\delta\left(\expval{ \sigma^z}{\psi_\xi(t)}-x\right)} = \Pclick_{\sigma^z}(x;t)+\Pnoclick_{\sigma^z}(x;t) \, , 
\end{equation}
i.e.\ the probability distribution of the expectation value of ${O} = \sigma^z$. It is easy to find that
$ \mathbb{T}(t) = \cos(J t)^2 \mathbb{I} + \sin(J t)^2 \sigma^x$,
%\begin{align}
%    \mathbb{T}(s'-s) = \begin{pmatrix*}[c]
%    \cos(J(s'-s))^2 && \sin(J(s'-s))^2\\
%    \sin(J(s'-s))^2 && \cos(J(s'-s))^2
%   \end{pmatrix*}, && \mathbb{V} = \frac{1}{\sqrt{2}} \begin{pmatrix*}
%        1 && 1 \\ 1 && -1
%    \end{pmatrix*},
%\end{align}
and $\mathbb{V} = (\sigma^z +\sigma^x)/\sqrt{2}$ whose columns are the eigenvectors of $\sigma^x$.
As expected $d_1=1$ and $d_2(t)=\cos(\Omega t)$, where $\Omega = 2J$ is the splitting between the energy eigenvalues of the system. Notice that, from these eigenvalues we can derive two recursion relations as in the integral equations~\eqref{eq:generic_recursion_relation}, then, since $d_1=1$ we immediately get $I_1(s)=e^{\gamma s}$. In the following, we will simplify the notation $I_2(s)=I(s)$. Finally, we notice that $\mathscr{D} (x,\sigma_n,t-s_n) =\delta \big(x-\sigma_n\cos(\Omega(t-s_n)) \big)$, from which we can compute the no-click contribution
\begin{equation}\label{eq:noclickspin}
    \Pnoclick_{\sigma^z}(x;t) = e^{-\gamma t} \, \delta \big(x-\cos(\Omega t) \big).
\end{equation} 
In order to find the full probability distribution, we can use equation~\eqref{eq:generic_click_2} to write 
\begin{equation}
\Pclick_{\sigma^z}(x;t) = \frac{e^{-\gamma t}}{2} \int_0^t \gamma \dd s \sum_{\sigma=\pm 1} \delta(x-{\sigma}\cos(\Omega(t-s))) \left[ e^{\gamma s} + \sigma I(s) \right] \label{eq:Pclick_single_spin}
\end{equation}
%where we simplified the notation $I_{n,2}(s)=I_n(s)$ and $\sum_n I_{n,2}(s) = I(s)$, with
%\begin{equation}
%    \begin{dcases}
%    I_{1}(s) = \cos(\Omega s),\\
%    I_{n}(s) = \int_0^{s}\gamma \dd s' \cos(\Omega(s-s')) I_{n-1}(s'),
%    \end{dcases}\label{eqs:SpinSystemRecursionRelation}
%\end{equation}
For a better understanding of the solution, we do not apply the Laplace method to solve the integral equation~\eqref{eq:volterra} for $I(s)$. Instead, we differentiate it twice with respect to $s$, which yields the following differential equation for $I(s)$\begin{equation}
\ddot{I} = \gamma \dot{I} - \Omega^2 I\\
\end{equation}
with the initial conditions $I(0) = 1$ and 
$\dot{I}(0) = \gamma$. This is the equation of motion of an ``anti-damped'' harmonic oscillator (since $\gamma>0$). The solution can be found easily using standard methods~\cite{morin_2008}, obtaining 
\begin{equation}
    I(s) = e^{\gamma s/2}\left[\cosh(\frac{\Gamma s}{2})+\frac{\gamma}{\Gamma}\sinh(\frac{\Gamma s}{2})\right], 
    %= e^{\gamma s/2}f(s),
\end{equation}
where we defined the following characteristic frequency $\Gamma = \sqrt{\gamma ^2-4 \Omega^2}$. As expected, we find three different regimes depending on the sign of $\gamma^2-4\Omega^2$. Let us analyze the behavior of the function $e^{-\gamma s/2}I(s)$. When $\gamma<2\Omega$, we have that $\Gamma$ is an imaginary quantity. Therefore, we find a regime in which the function oscillates with a frequency of $\Omega\sqrt{1-\gamma^2/(4\Omega^2)}$, which is lower than the natural frequency $\Omega$. When $\gamma = 2\Omega$, we get the critical regime for which the function $e^{-\gamma s/2}I(s)$ becomes linear, $\left(1+\gamma s/2\right)$. Finally, if $\gamma>2\Omega$, it grows exponentially with a rate $\gamma/2\sqrt{1-4\Omega^2/\gamma^2}$ smaller than $\gamma/2$.

%This means that the limit of $I(t)e^{-\gamma t}$ as $t$ approaches infinity is zero for all $\gamma>0$. Therefore we can write, for any value of $\gamma>0$, the expression of the asymptotic probability distribution in the case of $t\to\infty$ as
%\begin{equation}
%    P_{\sigma^z}(x) = \frac{\gamma}{2}\sum_{\sigma=\pm 1}\int_0^\infty\dd s\, \delta(x-\sigma\cos(\Omega s))e^{-\gamma s}
%\end{equation}

\begin{figure}
\centering
\begin{subfigure}{.33\textwidth}
  \centering
  \includegraphics[width=1.\linewidth]{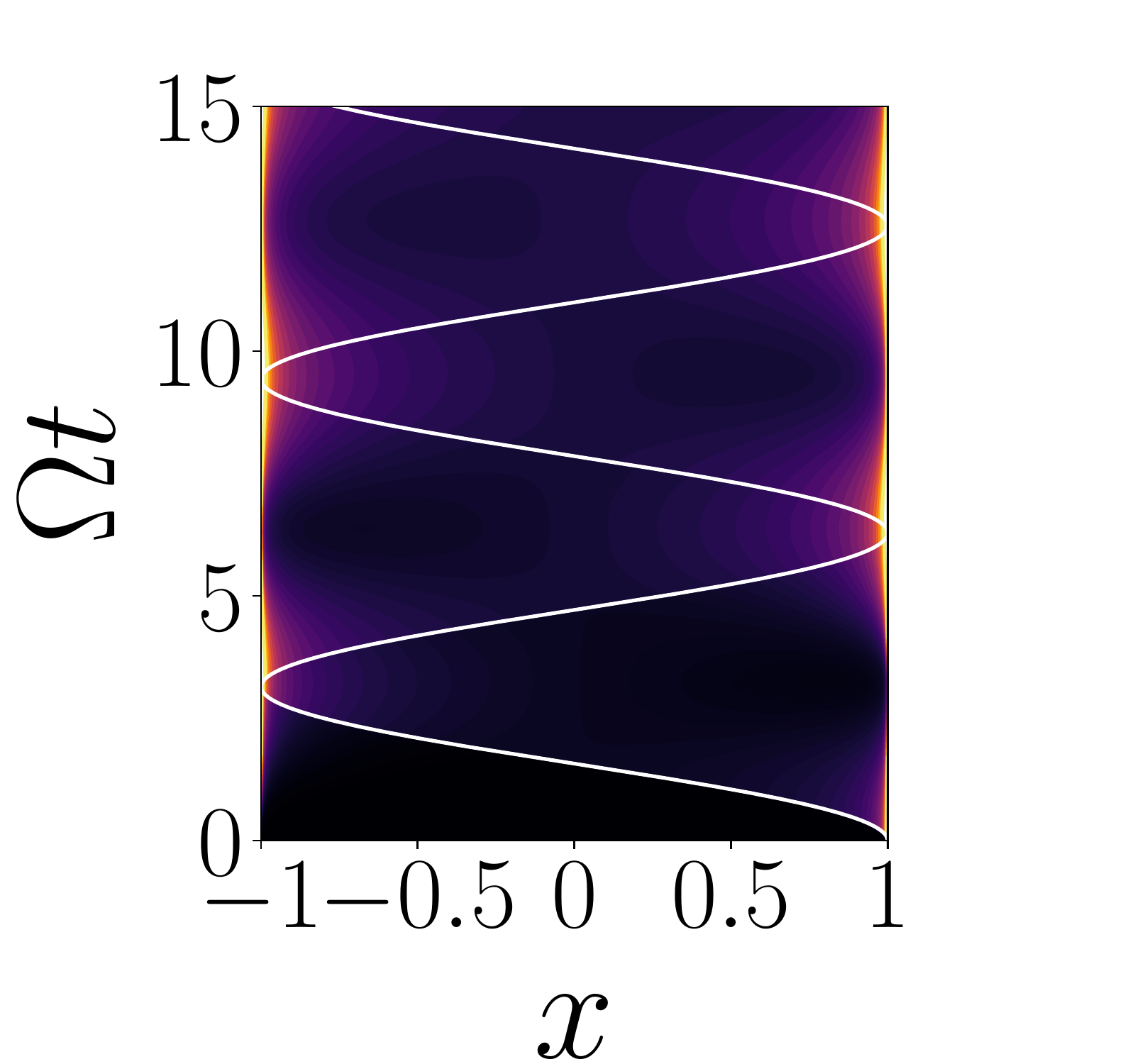}
  %\caption{A subfigure}
  \label{fig:sub1}
\end{subfigure}%
\begin{subfigure}{.33\textwidth}
  \centering
  \includegraphics[width=1.\linewidth]{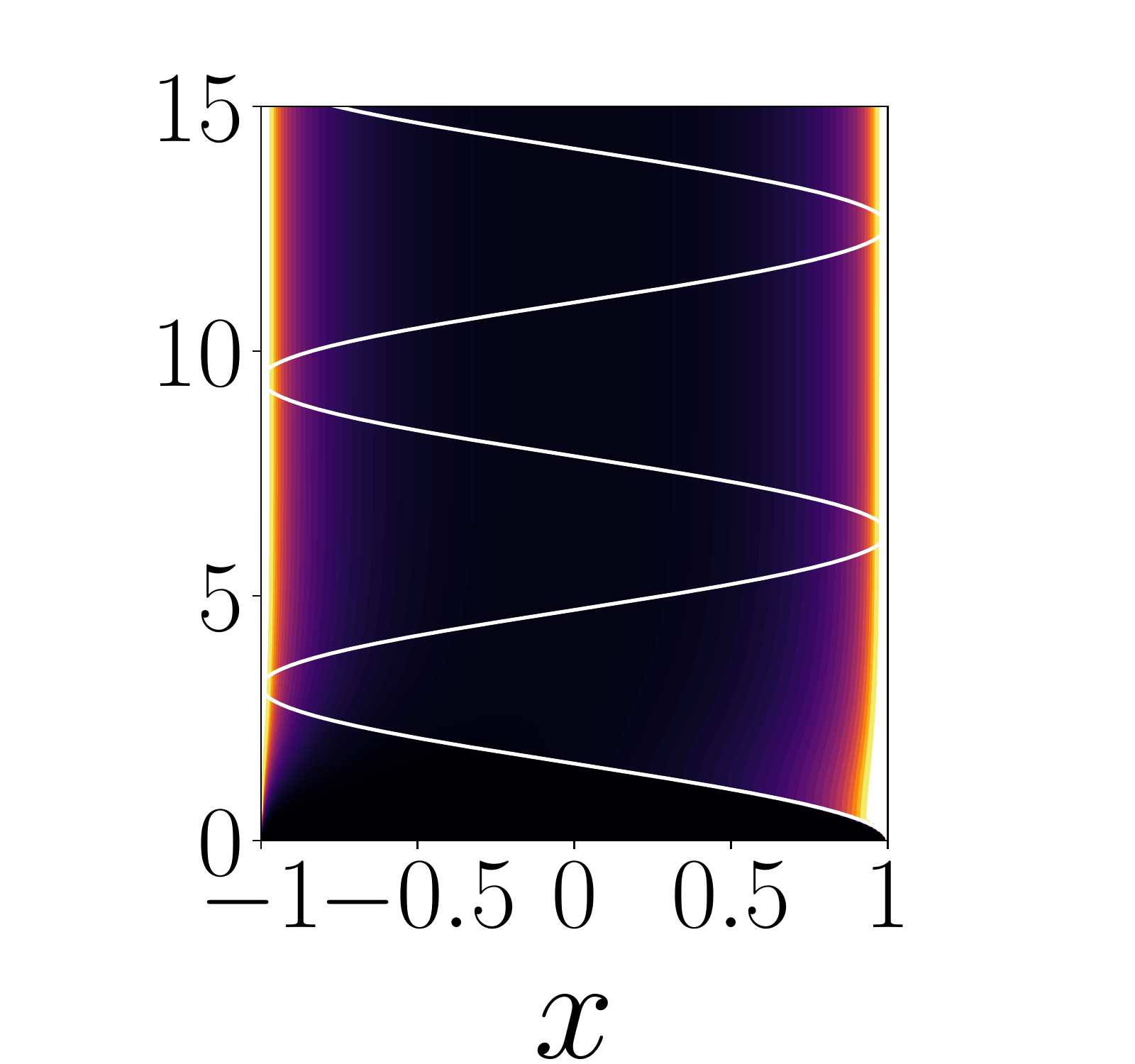}
  % \caption{A subfigure}
  \label{fig:sub2}
\end{subfigure}%
\begin{subfigure}{.33\textwidth}
  \centering
  \includegraphics[width=1.\linewidth]{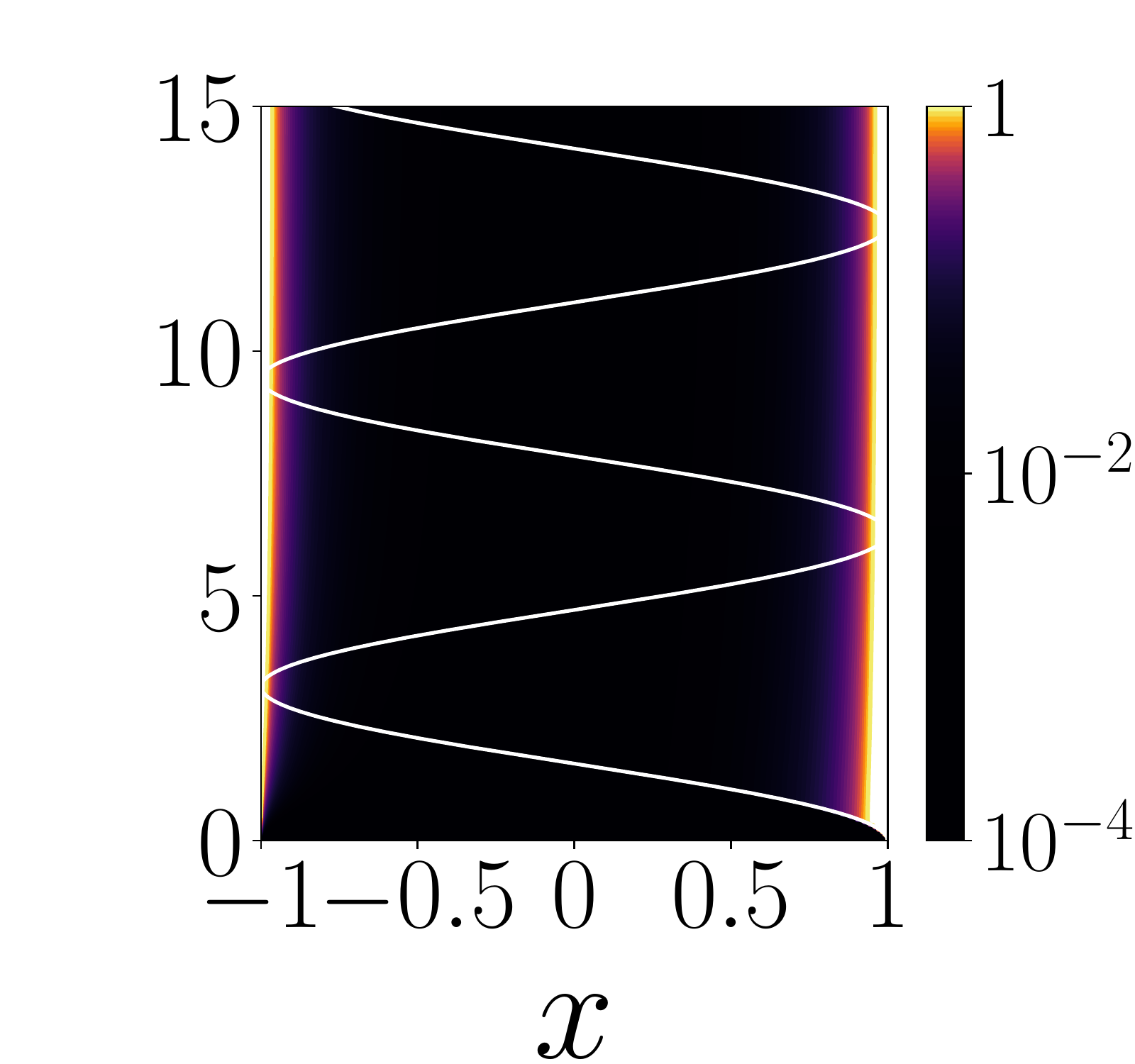}
  % \caption{A subfigure}
  \label{fig:sub3}
\end{subfigure} 
\caption{\label{fig:fig_1} The time-dependent probability distribution $P_{\sigma^z}(x;t)$ for a two-level system ($\Omega=1$). The white continuous line represents the no-click contribution $\Pnoclick(x;t)$ in Eq.~\eqref{eq:noclickspin}. 
We set $\gamma=0.2$ (left), $\gamma=2.0$ (center) and $\gamma=5.0$ (right).}
\label{fig:test}
\end{figure}

Let us finally consider the Dirac-delta function contribution. 
Solving for $x-\sigma \cos(\Omega(t-s)) = 0$, we easily get
\begin{equation}\label{eq:delta_poles}
    \tilde{s}_k = t \pm \frac{1}{\Omega}\arccos(x\sigma)+\frac{2\pi k}{\Omega},\quad k \in \mathbb{Z}.
\end{equation}
Using the properties of the Dirac-delta distribution, 
we can rewrite it as
\begin{equation}
    \delta(x-\sigma \cos(\Omega(t-s))) = \sum_{k \in \mathbb{Z}} \frac{\delta(s-\tilde{s}_k)}{\Omega\sqrt{1-x^2}}.
    \end{equation}
Since in Eq.~\eqref{eq:Pclick_single_spin}, the integral is taken over a finite interval $[0,t]$, we need to constraint the solutions $\tilde s_k$ in Eq.~\eqref{eq:delta_poles} only to those falling in that region. However, we may relax that condition by explicitly introducing the Heaviside step function $\theta(x)$, such that Eq.~\eqref{eq:Pclick_single_spin} finally reads
%\begin{equation}
%    \int_0^t \dd s \, \delta(g(s))f(s) = \sum_{k \in \mathbb{Z}}\frac{f(\tilde{s}_k)}{\Omega\sqrt{1-m^2}}\theta(t-\tilde{s}_k) \theta(\tilde{s}_k)
%\end{equation} 
%where we use $\theta(x)$ to denote the Heaviside step function. 
%By substituting this relation into Eq.~\eqref{eq:Pclick_single_spin}, we obtain
\begin{align}
\Pclick_{\sigma^z}(x;t) = \frac{\gamma e^{-\gamma t}}{2\Omega \sqrt{1-x^2}} \sum_{\sigma = \pm 1} \sum_{k \in \mathbb{Z}} \left[ e^{\gamma \tilde{s}_k} + \sigma I(\tilde{s}_k) \right]\theta(t-\tilde{s}_k) \theta(\tilde{s}_k) \, ,
\end{align}
where $\tilde s_k$ implicitly depends on $x$ and $t$ as well. In Figure~\ref{fig:test}, we plot the entire distribution $P_{\sigma^z}(x;t)$. As expected, at early time the probability is highly asymmetric, having support only in the vicinity of $x=1$. The no-click term is in fact localised at $x=\cos(\Omega t)$ but its weight is exponentially suppressed in time.
The click contribution is also showing a nontrivial asymmetric evolution. Its behavior strongly depends on $\gamma$ indeed: in the oscillatory regime ($\gamma < 2\Omega$), the probability distribution function bounces back and forth between the two extremes $x=\pm 1$ of its domain; after many oscillations, the number depending on the value of $\gamma$, it is expected to relax toward a symmetric distribution, the typical relaxation time being $\tau = 2/\gamma$. For $\gamma > 2\Omega$ the probability is not oscillating anymore and the relaxation time becomes $\tau = 2/(\gamma - \Gamma)$.
Interestingly, as the measurement rate is getting higher, the time needs for the magnetisation statistics to reach the equilibrium becomes larger and larger, diverging as $\tau \sim \gamma/\Omega^2$. 

\begin{figure}
    \centering
    \includegraphics[width=.45\linewidth]{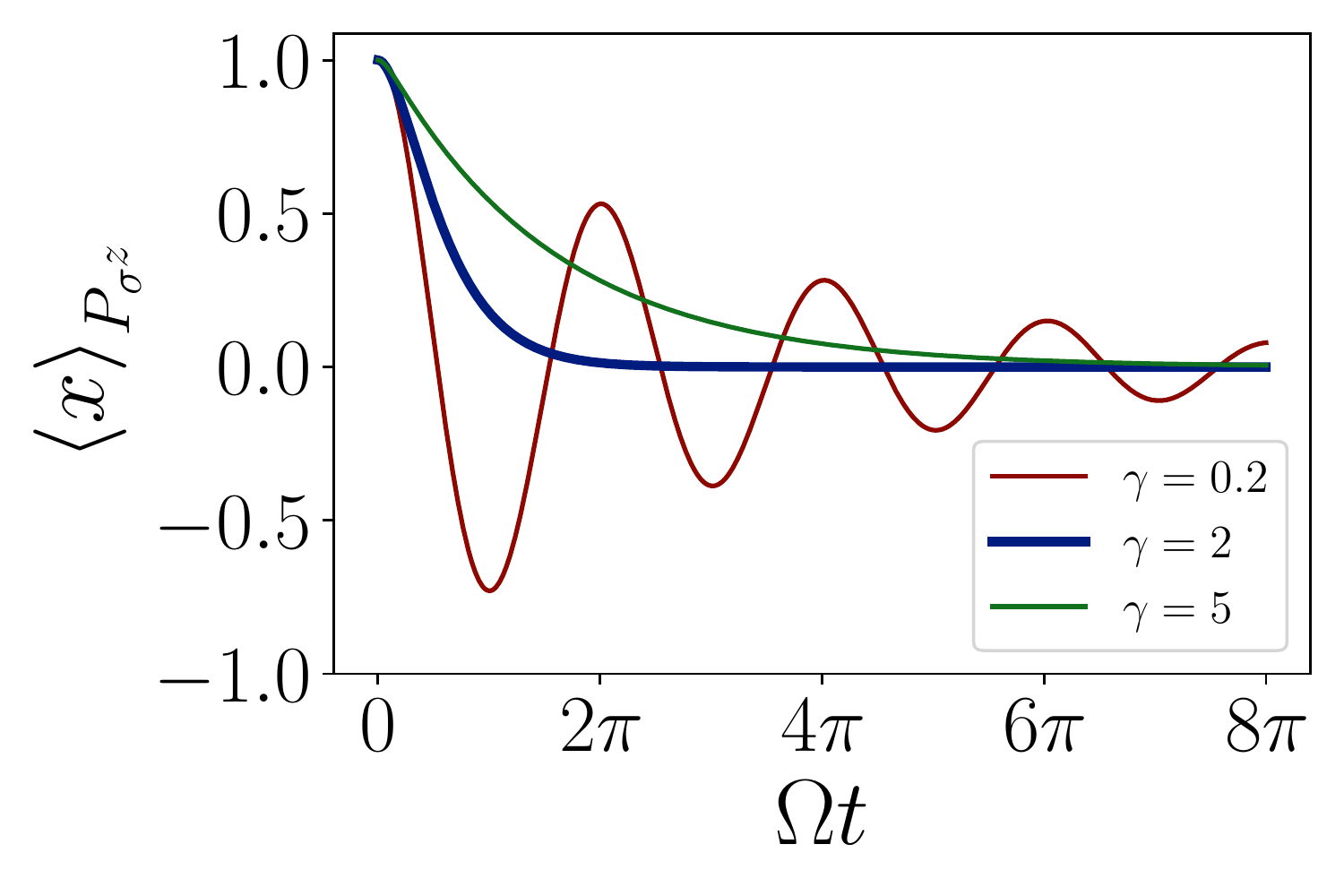}\qquad
    \includegraphics[width=.45\linewidth]{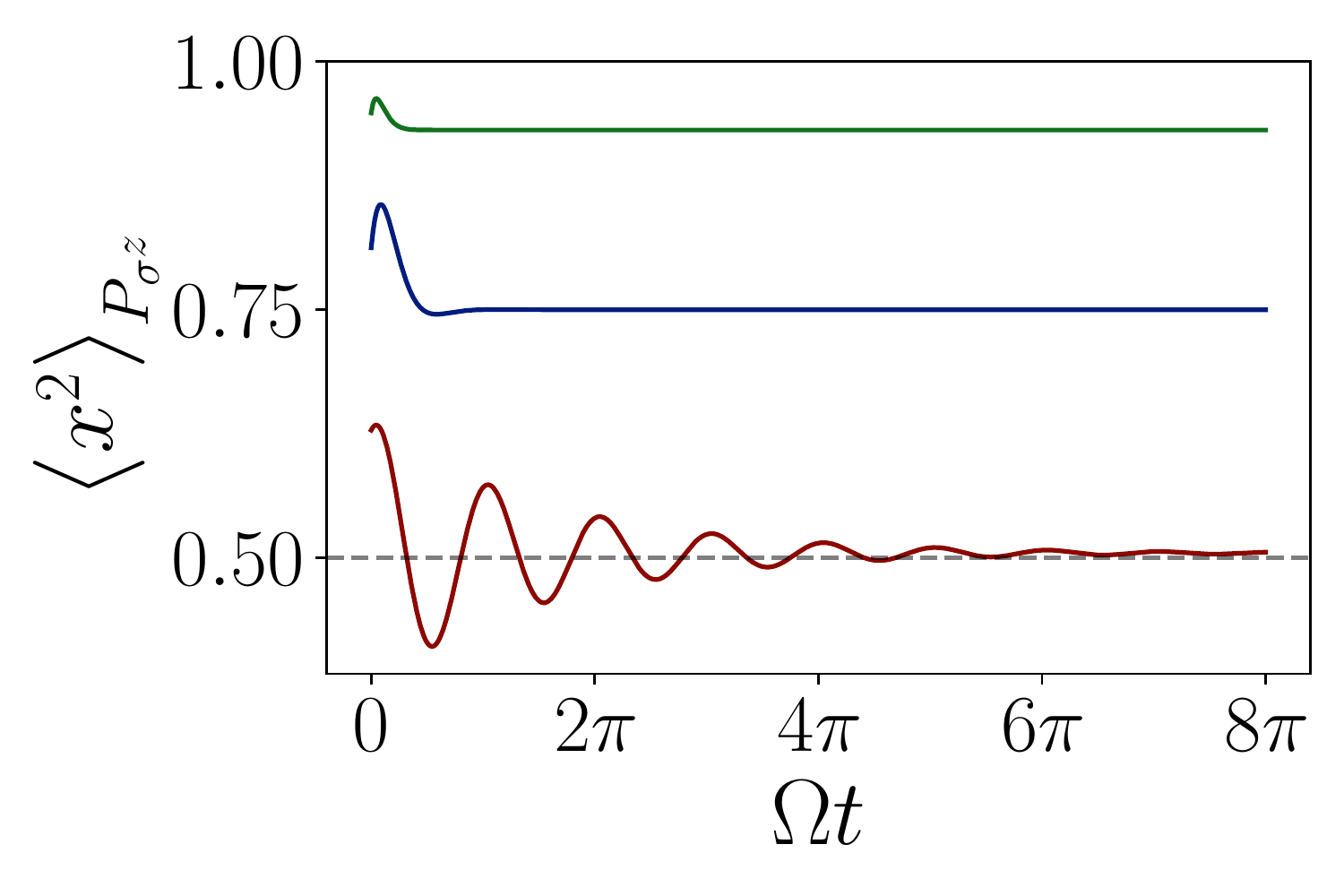}
    \caption{First (left) and second (right) moment of the probability distribution $P_{\sigma^z}(x;t)$ for a two-level system ($\Omega=1$). Notice that the critical value of $\gamma = 2$ leads to the fastest convergence to the equilibrium stationary value of the average magnetization. }
    \label{fig:first_second_moment_magnetization}
\end{figure}

The first moment of $P_{\sigma^z}(x;t)$, namely the magnetisation average $\moment{}{{\sigma^z}} = e^{-\gamma t} I(t)$,  does coincide with the expectation value over the averaged state (see~\ref{app:lindblad_spin}).
Nevertheless, our approach allows to easily compute the fluctuations of 
the magnetization along the trajectories. 
Indeed, by means of Eqs.~\eqref{eq:noclickspin} and \eqref{eq:Pclick_single_spin}, we can easily get the second moment of the distribution
\begin{equation}
    \moment{2}{{\sigma^z}} = e^{-\gamma t}\int_0^t \gamma \dd s \cos(\Omega(t-s))^2 e^{\gamma s} +  e^{-\gamma t}\cos(\Omega t)^2,
\end{equation}
which simplifies to
\begin{equation}
     \moment{2}{{\sigma^z}} = \frac{\gamma^2+2\Omega^2}{\gamma^2+4\Omega^2}+\frac{e^{-\gamma t}}{2(\gamma^2+4\Omega^2)}\left[2\gamma\Omega\sin(2\Omega t)+\Omega^2\cos(2\Omega t)\right],
\end{equation}
and which cannot be computed using the a Lindblad approach since it
corresponds to 
$\overline{\langle\psi_{\xi}(t)|\sigma_z |\psi_{\xi}(t) \rangle^2}$. In Figure~\ref{fig:first_second_moment_magnetization}, we represents $\moment{}{{\sigma^z}}, \moment{2}{{\sigma^z}}$ as a function of time for different values of the measurement rate $\gamma$. 

Finally, let us mention that one can easily obtain the asymptotic stationary distribution $P_{\sigma^z}(x,t\to\infty)$. Indeed, from a careful inspection of Eq.~\eqref{eq:Pclick_single_spin} one can argue that for large time the contribution coming from $I(s)$ is bounded by $e^{-\gamma t}\int_{0}^{t} \dd s | I(s) |$, thus decaying exponentially for large time. The only term that survives and contributes to the asymptotic stationary distribution is the one depending on $e^{\gamma s}$.
Therefore, for any finite $\gamma$, the stationary distribution can be exactly evaluated as
\begin{equation*}
P_{\sigma^z}(x) = \frac{\gamma}{2}\sum_{\sigma=\pm 1}\int_0^\infty\dd s\, \delta(x-\sigma\cos(\Omega s))e^{-\gamma s} = \frac{\gamma \left[e^{2\,\gamma \arcsin(x)/\Omega}+1\right] e^{\gamma \arccos(x)/\Omega}}{2 \Omega \left(e^{\pi  \gamma/\Omega}-1\right) \sqrt{1-x^2}},
\end{equation*}
which, as expected, is an even function of $x$. 
Interestingly, all (non-vanishing) stationary moments can be easily computed from the integral representation of the probability, namely
\be
\moment{2n}{{\sigma^z}} =
\int_{0}^{\infty} ds \, e^{-s} \cos^{2n}(s \Omega/\gamma) =
\frac{_2 F_1(-2n,-n-i \gamma/(2\Omega);1-n-i \gamma/(2\Omega);-1)}{2^{2n} (1 - 2 i n \Omega/\gamma)},
\ee
while the odd moments are identically vanishing.
In particular, when $\gamma\to\infty$, $\moment{2n}{{\sigma^z}}\to 1$ for all $n$, confirming the fact that the stationary distribution converges to $[\delta(x-1)+\delta(x+1)]/2$.
% Let us stress that the entire probability distribution is much more informative than the simple mean state whose dynamics is described by the Lindblad equation.

\section{Hopping particle}
A free hopping quantum particle propagates in a lattice with a ballistic spreading. However, there are ways to prevent or slow down the propagation as, for instance, adding a disorder potential which induces Anderson localization \cite{AndersonLocalization1958,scardicchio2017perturbation}. Here, we show that the quantum Zeno effect due to the coupling of the hopping particle to a measurement apparatus can also results into a slowdown of the particle propagation \cite{khosla2021quantum,Das2022JStat}. Related protocols have been studied in the context of quantum stochastic resetting, in which the hopping particle is reset to the initial state with a certain probability \cite{Majumdar2018PRB}.

We consider a simple hopping fermion on a 1D lattice, whose hamiltonian reads
\begin{equation}
     H = -\Omega \sum_{j=1}^{L} \bigg( c^\dagger_j c_{j+1} + c^\dagger_{j+1} c_j \bigg)
\end{equation}
with periodic boundary conditions (i.e. $c_{j+L} = c_j$).
Here the lattice dimension $L$ (even) plays the role of an infrared cutoff.
In the following we will take the limit $L\to\infty$ whenever it will be unambiguous.  
The Fermionic operator satisfy the canonical anti-commutation relations $\{c_i,c^{\dag}_j\}=\delta_{ij}$. We define the Fourier modes operators
\begin{equation}
    \tilde c_{k} =  \frac{1}{\sqrt{L}}\sum_j e^{ik j} c_j, \quad
    c_j =\frac{1}{\sqrt{L}} \sum_k e^{-ikj}  \tilde c_k,
\end{equation}
such that $\{\tilde c_p,\tilde c^{\dag}_q\}=\delta_{pq}$, and $k \in \{-\pi,-\pi+2\pi/L,-\pi+4\pi/L,\dots, \pi\}$.
The Hamiltonian become diagonal in the Fourier representation, i.e.
$
H =  \sum_k \varepsilon(k)  \tilde c^\dagger_k  \tilde c_k
$,
where $\varepsilon(k) = -2 \Omega \cos k$. 
We now restrict the problem to the single-particle sector of the hamiltonian, we can define the states $\ket{j} = c^\dagger_j\ket{\emptyset}$ and $\ket*{\tilde k} = \tilde c^\dagger_k\ket{\emptyset}$, which represents the particle in position $j$ or with momentum $k$ respectively. Notice that $\braket*{j}{\tilde k}=\exp(-ikj)/\sqrt{L}$ is the normalised wave function. 
Since $H$ commute with he total number of particles, the unitary dynamics can be restricted in such sector and it is governed by the following single-particle Hamiltonian
\begin{equation}\label{eq:hoppins_ham}
     H = -\Omega\sum_{j=1}^{L} \left( \dyad{j}{j+1} + \dyad{j+1}{j} \right) = \sum_k \varepsilon(k)\dyad*{\tilde k}.
\end{equation}

\begin{figure}
\centering
\begin{subfigure}{.33\textwidth}
  \centering
  \includegraphics[width=1.\linewidth]{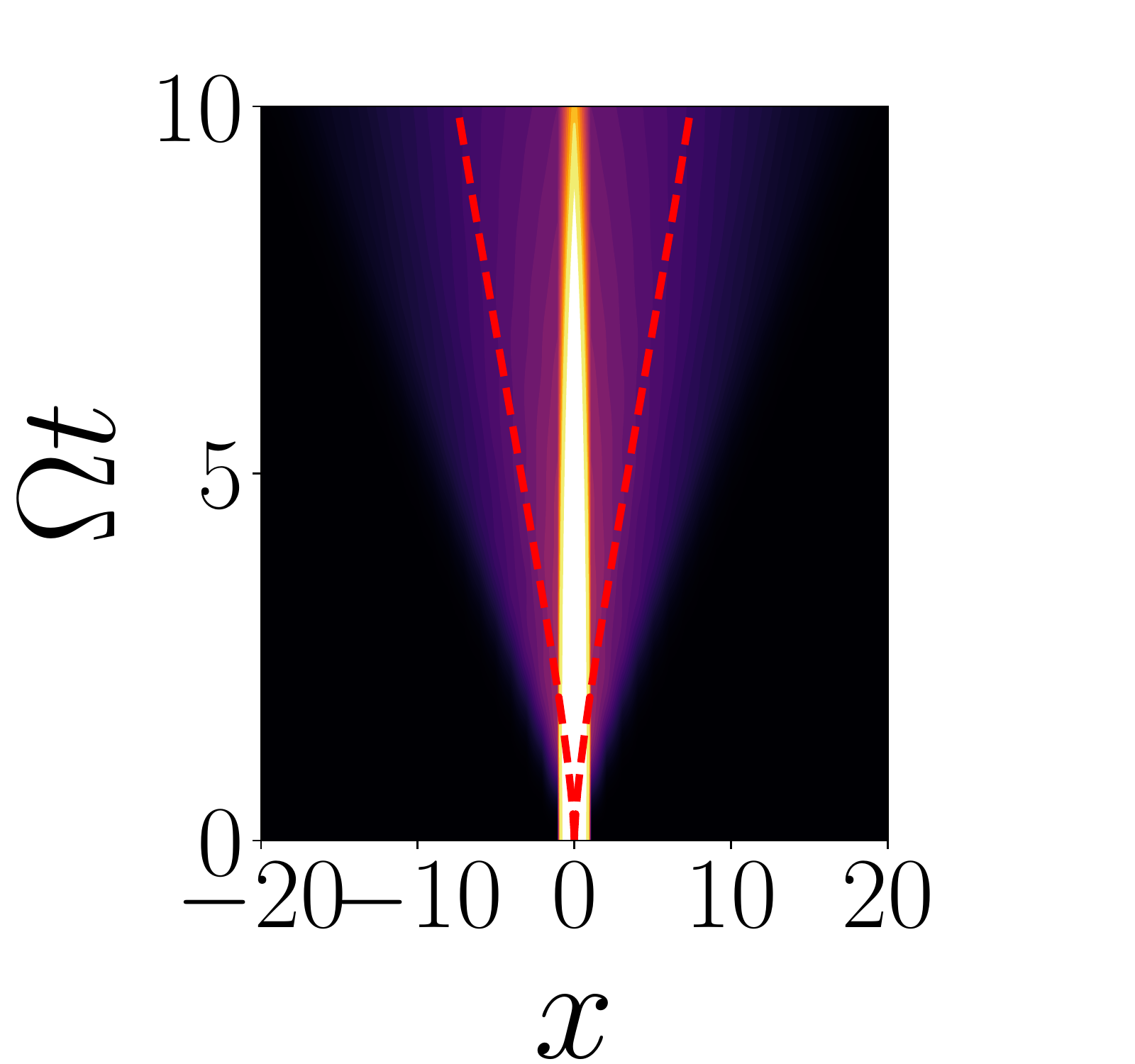}
  %\caption{A subfigure}
  \label{fig:subh1}
\end{subfigure}%
\begin{subfigure}{.33\textwidth}
  \centering
  \includegraphics[width=1.\linewidth]{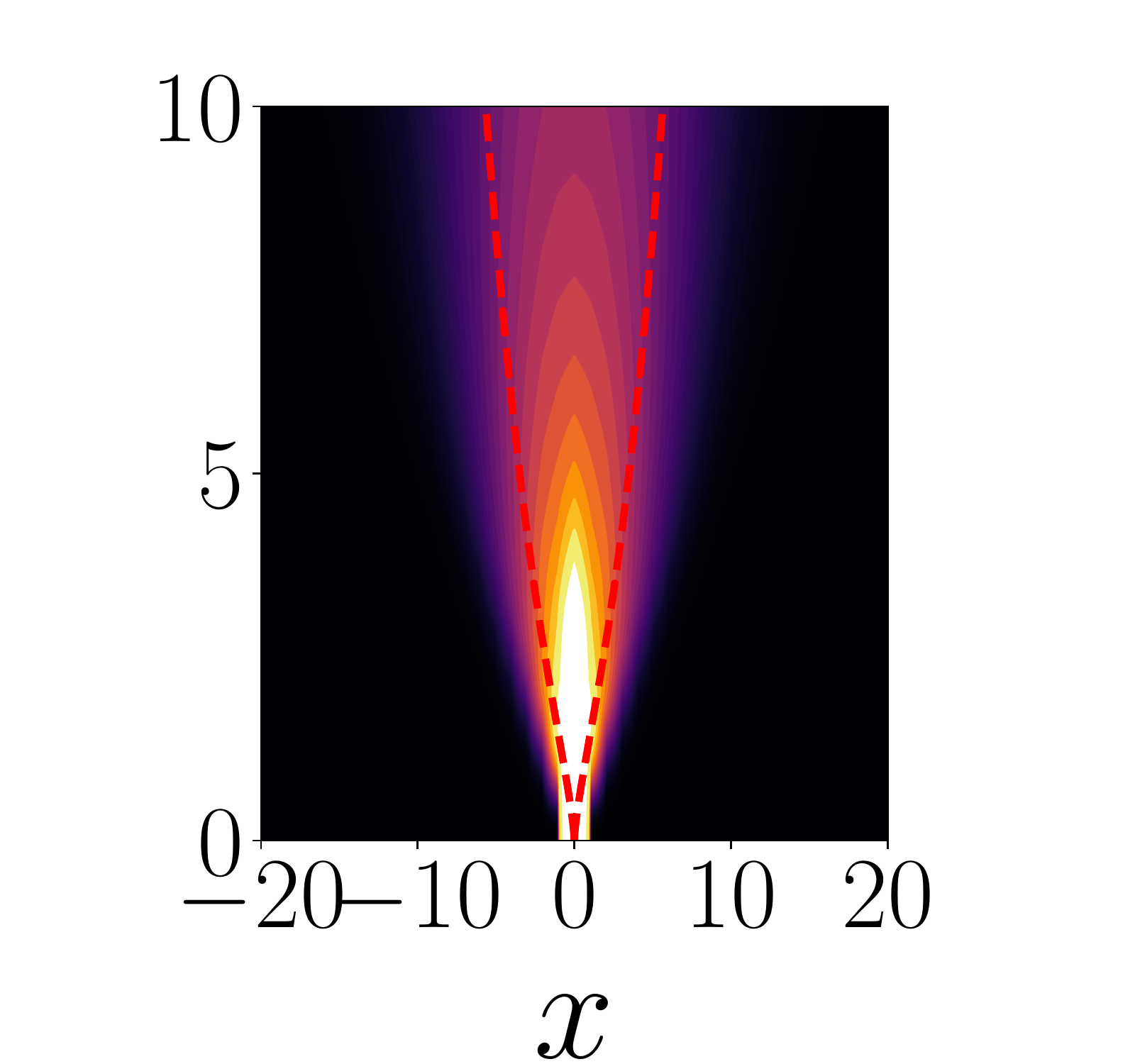}
  % \caption{A subfigure}
  \label{fig:subh2}
\end{subfigure}%
\begin{subfigure}{.33\textwidth}
  \centering
  \includegraphics[width=1.\linewidth]{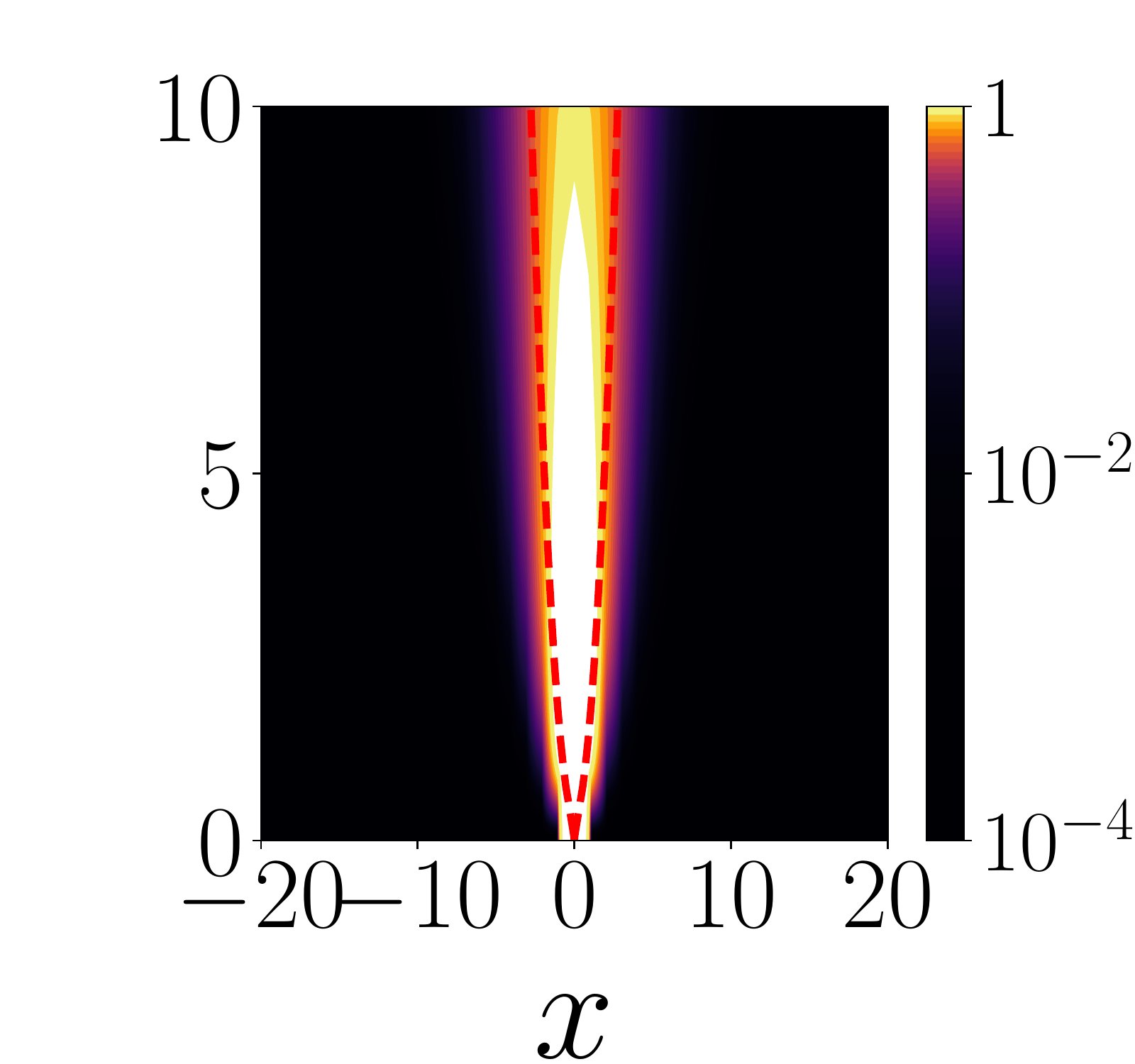}
  % \caption{A subfigure}
  \label{fig:subh3}
\end{subfigure}
\begin{subfigure}{.33\textwidth}
  \centering
  \includegraphics[width=1.\linewidth]{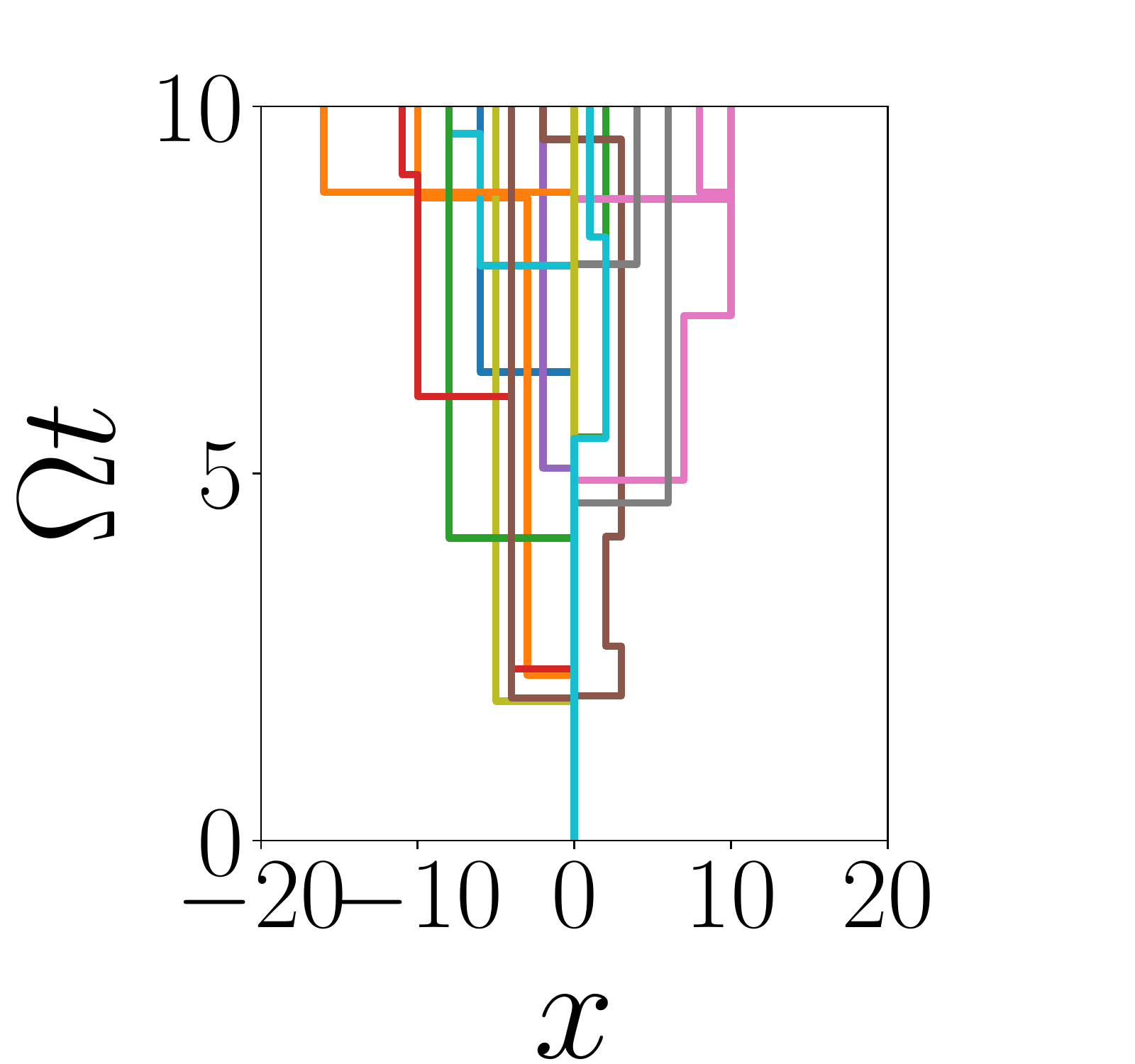}
  %\caption{A subfigure}
  \label{fig:subh12}
\end{subfigure}%
\begin{subfigure}{.33\textwidth}
  \centering
  \includegraphics[width=1.\linewidth]{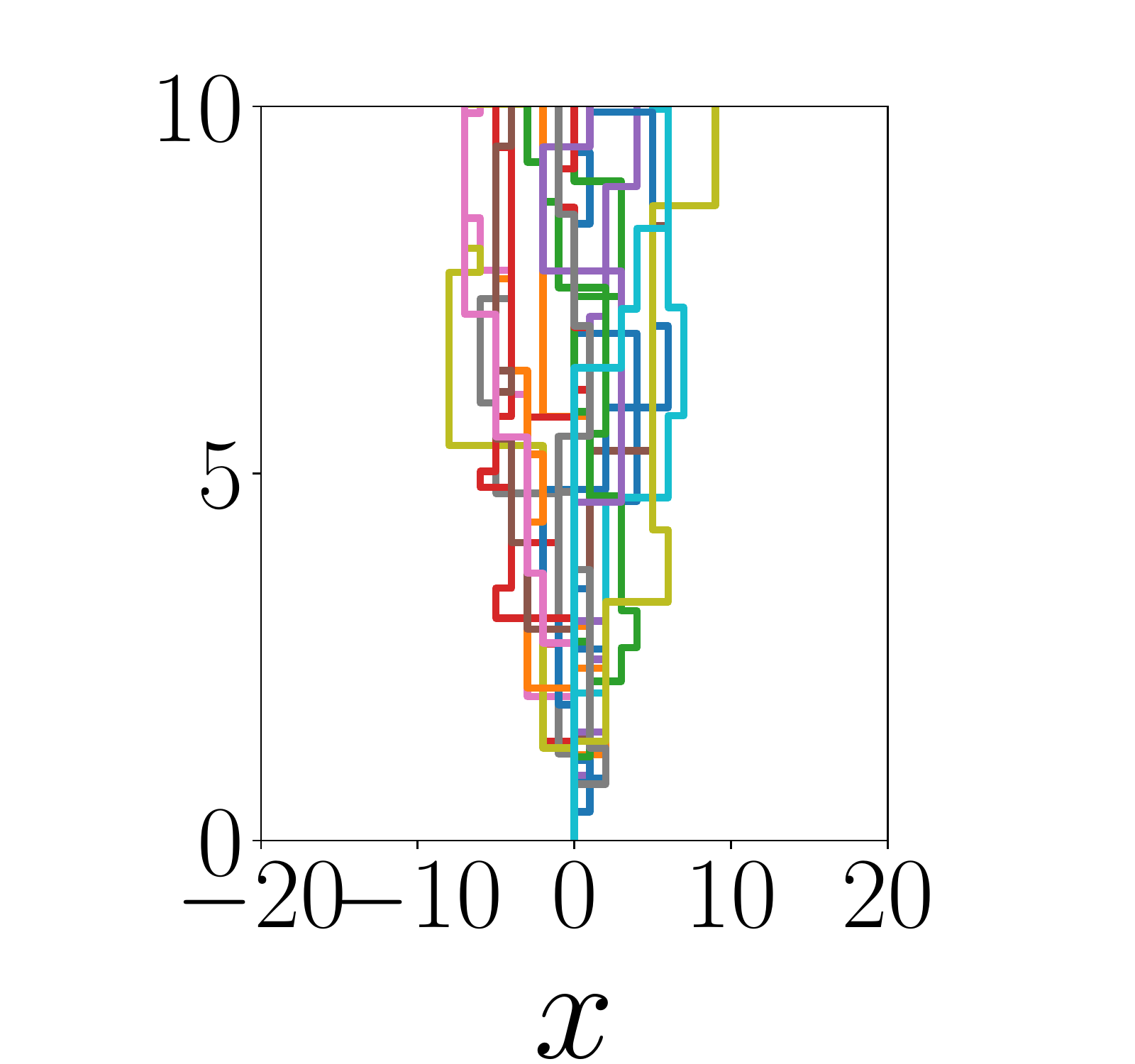}
  % \caption{A subfigure}
  \label{fig:subh22}
\end{subfigure}%
\begin{subfigure}{.33\textwidth}
  \centering
  \includegraphics[width=1.\linewidth]{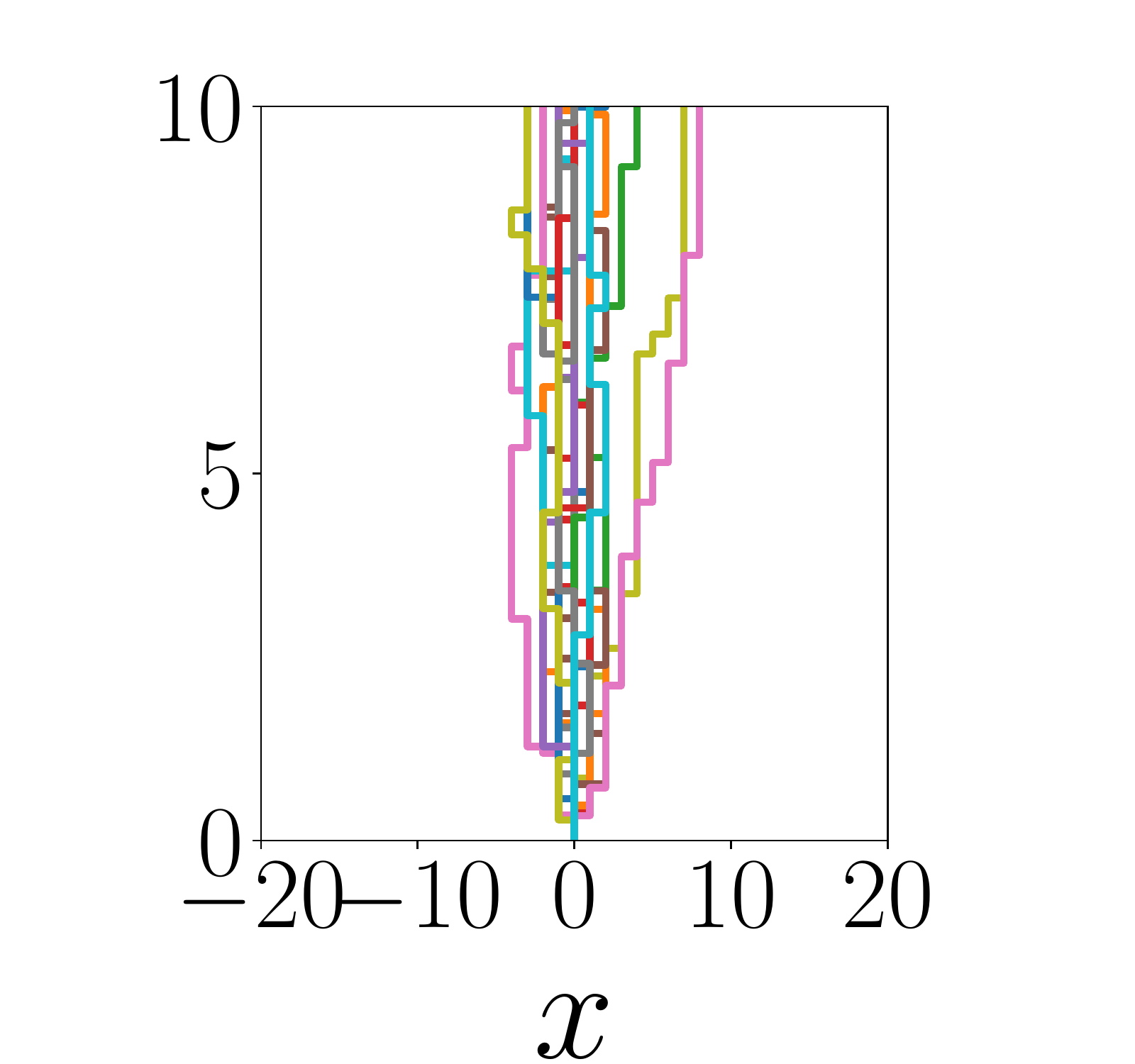}
  % \caption{A subfigure}
  \label{fig:subh32}
\end{subfigure}
\caption{\label{fig:fig_2} Upper panels: the time-dependent probability distribution $P_q(x;t)$ for an hopping particle ($\Omega=1$). Red lines represents the standard deviation $\pm \moment{2}{q}^{1/2}$ as in Eq.~\eqref{eq:momento2hopping}. 
We set $\gamma=0.25$ (left), $\gamma=1.0$ (center) and $\gamma=5.0$ (right). Lower panels: the expectation value $\expval{q}{\psi_\xi(t)}$ for $20$ trajectories and the same values of $\gamma$.}
\label{fig:test_hopping}
\end{figure}

We consider the particle initial localised at the origin $j=0$ of our lattice, namely $\ket{\psi_{\xi}(0)} = \ket{0}$ for all trajectories $\xi$. We then suppose to continuously measure, with a rate $\gamma$, the position operator 
\begin{equation}
q = \sum_{j} j \dyad*{j}. 
\end{equation}
We are thus interested in the displacement of the particle along each single trajectory, however, for symmetry reasons, 
when no measurement occurs (i.e. $\gamma=0$)
the probability function of the outcome of $\langle q\rangle_t$ is time independent, namely 
$\overline{\delta\left(x-\langle q\rangle_t\right)} = \delta(0)$.
This is not in contradiction with the expected ballistic spreading under the free evolution, which can be extracted when observing even power of $q$ (see~\ref{app:moment_qm}).
Notice that, this is not true anymore when $\gamma\neq 0$. We are thus interested in the probability distribution function of the particle displacement itself, namely \begin{equation}
    P_q(x;t) = \overline{\delta\left(\expval{q}{\psi_\xi(t)}-x\right)} 
    =
    \Pclick_q(x;t)+\Pnoclick_q(x;t),
\end{equation}
where in this case the no-click contribution is trivially given by
$\Pnoclick_q(x;t) = e^{-\gamma t} \delta(x)$, 
and we used the following results for the evolution amplitudes in the thermodynamic limit
\begin{align}
    \mel{j}{U(t)}{l} = \sum_{k} \mel*{j}{e^{-i t H}}{\tilde k}\braket*{\tilde k}{l} &= \int_{-\pi}^\pi \frac{\dd k}{2\pi} e^{i\left[k(j-l)-2t\Omega\cos k\right]}\label{eq:matrix_element_hopping} &= (-i)^{j-l}J_{j-l}(2\Omega t),
\end{align}
with $J_n(x)$ being the Bessel function of the first kind. Now the transition probability matrix reads
$\mathbb{T}_{i,j}(t) = J^2_{i-j}(2\Omega t)$,
which is a circulant matrix due to translational invariance. Let us define the not normalised eigenvectors whose component are $\mathbb{V}_{n,k} = e^{-ink}$, such that \begin{equation}
    (\mathbb{T}(t)\mathbb{V})_{n,k}  =\sum_j J^2_{n-j}(2\Omega t) e^{-ijk} = e^{-ink}\sum_l J^2_j(2\Omega t)e^{ilk} \equiv \mathbb{V}_{n,k} \, d_k(t) 
\end{equation}
where we identified the eigenvalues of the transfer matrix as
%\begin{equation}
$d_k(t) = \sum_l J^2_l(2\Omega t)e^{ikl}
= J_{0}[\omega_k t]$,
%\end{equation}
with $\omega_k = 4 \Omega \sin(k/2)$. 
Following Section~\ref{sec:protocol}, we can easily
 solve the integral equation~\eqref{eq:volterra}
for $I_k(s)=\sum_{n=1}^\infty I_{n,k}(s)$ with kernel $J_{0}[\omega_k t]$; the Laplace transform reads
 %easily get the following recursion relation
%\begin{equation}
%    \begin{dcases}
%    I_{1,k}(s) = J_{0}[\omega_k s],\\
%    I_{n,k}(s) = \int_0^{s}\gamma \dd s' 
%    J_{0}[\omega_k (s - s')]  I_{n-1,k}(s'),
%\end{dcases}\label{eqs:HoppingRecursionRelation}
%\end{equation}
%\begin{equation}
%    I_k(s) - J_0(\omega_k s) = \int_0^s \gamma \dd s' J_0(\omega_k(s-s'))I_k(s').\label{eq:integral_equation_hopping}
%\end{equation}
\begin{equation}
    \mathscr{L}[I_k](z) = \frac{1}{\sqrt{z^2+\omega_k^2}-\gamma}
    =\sum_{n=1}^{\infty}\frac{\gamma^{n-1}}{(z^2+\omega_k^2)^{n/2}},
\end{equation}
and we can identify with $\mathscr{L}[I_{n,k}](z)$ the terms of the series expansion, thus finally getting 
\begin{equation}
    I_{n,k}(s) = \sqrt{\pi } \gamma ^{n-1}\left(\frac{s}{2 |\omega_k|}\right)^{\frac{n-1}{2}}\frac{  J_{\frac{n-1}{2}}(s |\omega_k|)}{\Gamma \left(n/2\right)}.
\end{equation}
%Let us now consider the delta function\begin{equation}
%    \mathscr{D} (x_2,j_n,t-s_n)=\delta(x_2-\mel{j_n}{{U}_{t,s_n}^\dagger q^2{U}_{t,s_n}}{j_n}) 
%\end{equation}
%we get that\begin{align}
%    \mel{j_n}{{U}_{t,s_n}^\dagger q^2{U}_{t,s_n}}{j_n}&= \sum_j j^2  \mel{j_n}{{U}_{t,s_n}^\dagger \dyad{j}  {U}_{t,s_n}}{j_n}\\ \notag&= \sum_j j^2 J_{j-j_n}^2(2(t-s_n)) = 2(t-s_n)^2+j^{2}_n
%\end{align}
With a simple generalization of equation~\eqref{eq:generic_click}, we can write the click contribution to the probability distribution as
\begin{equation}\label{eq:Pclick_hopping}
     \Pclick_{q}(x;t) = e^{-\gamma t} \sum_{j} \int_0^t \gamma \dd s 
     \int_{-\pi}^\pi \frac{\dd k}{2\pi} 
     \mathscr{D} (x,j,t-s)  e^{-ijk} I_{k}(s),
\end{equation}
where, thanks to the properties of the Bessel functions, the delta contribution reduces to
$
\mathscr{D} (x,j,t-s)
%=\delta(x-\mel{j}{{U}_{t,s}^\dagger q{U}_{t,s}}{j})  
= \delta(x-j)$,
which basically says that the probability has only support on $x\in\mathbb{Z}$. The entire probability distribution $P_q(x;t)$ is represented in Figure~\ref{fig:fig_2} together with some representative quantum trajectories for different values of $\gamma$. 

We have previously observed that the first moment of the probability distribution is identically vanishing due to the inversion symmetry. The second moment instead gets a nontrivial contribution from the $\Pclick_q$ part, thus reading \begin{equation}
    \moment{2}{q} = e^{-\gamma t}\int_0^t \gamma \dd s\, \int_{-\pi}^\pi \frac{\dd k}{2\pi} \left[\sum_j j^2 e^{-ij k}\right]I_k(s),
\end{equation}
which can be further simplified using the identity
$
\sum_j j^2 e^{-ijk} = -\sum_j \partial_k^2 e^{ijk} = - 2 \pi \delta''(k),
$
leading to
\begin{equation}\label{eq:momento2hopping}
    \moment{2}{q} = -e^{-\gamma t }\int_0^t \gamma \dd s \, [\partial^2_k I_k(s)]\bigr\rvert_{k=0} = \frac{4\Omega^2}{\gamma^2} \left[(\gamma  t-2)+e^{-\gamma  t} (\gamma  t+2)\right].
\end{equation}
The second moment does behave differently depending on the time-scale, 
showing the following scaling behaviour \begin{equation}\label{eq:x2_expansion}
    \moment{2}{q} \sim \begin{dcases}
        \gamma \Omega^2t^3\quad \gamma t \ll 1\\
        \frac{4\Omega^2t}{\gamma}
 \quad \gamma t \gg 1
    \end{dcases}.
\end{equation}

Let us mention that the fluctuations of the $P_q(x;t)$ distribution cannot give information about the ballistic behavior at $\gamma=0$, due to the inversion symmetry. As a matter of fact, what Eq.~\eqref{eq:x2_expansion} gives us is the leading term for $\gamma\neq 0$. In the asymptotic regime $\gamma t \gg 1$, it does coincide (as expected) with $\moment{}{q^2}$ confirming the diffusive behavior for any finite measurement rate. However, in the $\gamma t \ll 1$ regime, the $O(\gamma^0)$ term is missing, and it is recovered in the expansion of $\moment{}{q^2}$ (see~\ref{app:moment_qm}).

\section{Conclusions}
In this article, we have presented an approach for studying the dynamics of quantum systems undergoing unitary evolution and continuous monitoring. Our approach goes beyond the traditional Lindblad master equation and provides a more complete picture of the system's behavior by considering the entire ensemble of stochastic quantum trajectories.

Specifically, we have developed an analytical tool to compute the probability distribution of the expectation value of a given observable over the ensemble of quantum trajectories. We obtained exact formulas to evaluate this probability distribution and its moments for two paradigmatic examples: a single qubit subjected to magnetization measurements, and a free hopping particle subjected to position measurements. 

Our results demonstrate that the probability distribution of expectation values can exhibit non-trivial features that are missed by traditional approaches, highlighting the full properties contained in the set of quantum trajectories in the analysis of continuously monitored quantum systems.
\newline

\noindent \textit{Note added.}---While completing this work, the
preprint \cite{Romito2023Related} appeared, dealing with topics related to the ones discussed by us.

\subsection*{Acknowledgments}
This work was supported by the PNRR MUR project PE0000023-NQSTI

\newpage
\appendix

\section{Useful properties of the Bessel Functions}
Here we collect some properties of the Bessel functions\begin{equation}
    J_n(t)=\frac{1}{2\pi} \int_{-\pi}^\pi \dd z\, e^{i\left(nz -t\sin z\right)},
\end{equation}
which have been used all along the main text.
Let's start by noticing that $J_{n}(t)^2 $
are non-negative and normalized over $\mathbb{Z}$, i.e. $\sum_{n\in\mathbb{Z}}J_{n}(t)^{2} = 1$.
In other words, they can be interpreted as probabilities over the infinite set of discrete events $n\in\mathbb{Z}$, with a parameter dependence $t$. They indeed satisfy the following property
\begin{equation}
    \sum_{n\in\mathbb{Z}} x^{2n} J_{n}(t)^2 = 
J_{0}[i t (x-1/x)],
\end{equation}
from which we can easily define the generating function $F_t(\lambda)$ of the moments of the distribution, as 
\begin{equation}
    F_{t}(\lambda) \equiv \sum_{n\in\mathbb{Z}}
    e^{i\lambda n}J_{n}^{2}(t) =
    J_{0}[2t \sin(\lambda/2)],
\end{equation}
such that $\sum_{n\in\mathbb{Z}} n^{k} J_{n}(t)^2 = (-i)^{k}\partial^{k}_{\lambda}F_t(\lambda)$.
From the previous relation it is straightforward to show that
\be
\int_{-\pi}^{\pi}\frac{\dd \lambda}{2\pi} e^{-i \lambda n}
 J_{0}[2t \sin(\lambda/2)]
 = J^2_n(t)
\ee

Finally, another useful property of the Bessel functions is that it is possible to compute the Laplace Transform, which reads
\begin{equation}
    \mathscr{L}[J_n](z) = 
    \frac{\left(z+\sqrt{z^2+1}\right)^{-n}}{\sqrt{z^2+1}}.
\end{equation}

\section{Lindblad equation solution for the two level system}\label{app:lindblad_spin}

When we are interested in the dynamical map averaged over the quantum trajectories, the measurement protocol outlined in the main text can be reformulated in terms of a Lindblad equation for the averaged density matrix $\rho(t)=\overline{\dyad{\psi_{\xi}(t)}}$.
In fact, averaging over different trajectories does correspond to relax both the information on whether the spin has been measured, and the result of the measurement itself. { See \cite{Barthel2013PRL} for some results of projective measurement-based dissipation descriptions of Lindblad equations for quantum spin systems.}

In particular for a single spin undergoing projective measurements of $\sigma^{z}$ (see Sec.~\ref{sec:single_spin}), the average state $\rho$ transforms accordingly to 
\be
\rho(t) \to \left( 1-\frac{\gamma dt}{2} \right)
 \rho(t) + \frac{\gamma dt}{2}  \sigma^{z} \,  \rho(t)
\,  \sigma^{z},
\ee
where $\gamma dt$ is the probability that the system is measured, after a discretization of the continuum time evolution has been applied. 

Combining the previous expansion with the unitary part of the evolution, and taking the continuum limit $dt\to0$ with $\gamma$ fixed, we finally get the following Lindblad master equation 
\be\label{eq:lindblad}
\partial_t  \rho = - i [ H, \rho]
+ \frac{\gamma}{2}
\left(  \sigma^{z} \,  \rho \,  \sigma^{z}
- \rho \right),
\ee
with $ H = -J  \sigma^{x}$. This equation can be easily solved by expanding the density operator in the basis of Pauli matrices
\be
\rho(t) =\frac{1}{2}\mathbb{I} + \frac{1}{2}\sum_{\alpha=1}^3 m_{\alpha}(t) \sigma^{\alpha},
\ee
where $m_{\alpha}(t) = \Tr[\sigma^{\alpha}\rho(t)]$ and $\mathbb{I}$ is the $2\times 2$ identity matrix. 
The Lindblad equation becomes a linear differential equation
for the three components of the magnetisation $(m_x,m_y,m_z)^T$, which reads
\be
\partial_t
\begin{pmatrix}
m_{x}\\
m_{y}\\
m_z
\end{pmatrix}
= 
\begin{pmatrix}
-\gamma & 0 & 0\\
0 & -\gamma & \Omega\\
0 & -\Omega & 0
\end{pmatrix}
\begin{pmatrix}
m_{x}\\
m_{y}\\
m_z
\end{pmatrix}
\ee
where $\Omega = 2J$. The $z$ component evolves following the differential equation of a damped harmonic oscillator  
$
\partial^2_t m_z = - \Omega^2 m_z -\gamma \partial_t m_z
$,
with initial condition $m_z(0) = 1$, $\partial_t m_z(0) = 0$.
From the solution of such equation we easily recover the result  of the main text, namely
\be
m_z(t) =  e^{-\gamma t /2}
\left[\cosh(\frac{\Gamma t}{2})+\frac{\gamma}{\Gamma}\sinh(\frac{\Gamma t}{2})\right] \equiv \moment{}{\sigma^z}
\ee
with $\Gamma = \sqrt{\gamma ^2-4 \Omega^2 }$. In addition, we also gets
\be
m_x(t) = 0, \quad 
m_y(t) = 
 e^{-\gamma t /2}
\frac{2\Omega}{\Gamma}\sinh(\frac{\Gamma t}{2}).
\ee

\section{Lindblad equation solution for the hopping particle}\label{app:moment_qm}
\begin{figure}
    \centering
    \includegraphics[width=.7\linewidth]{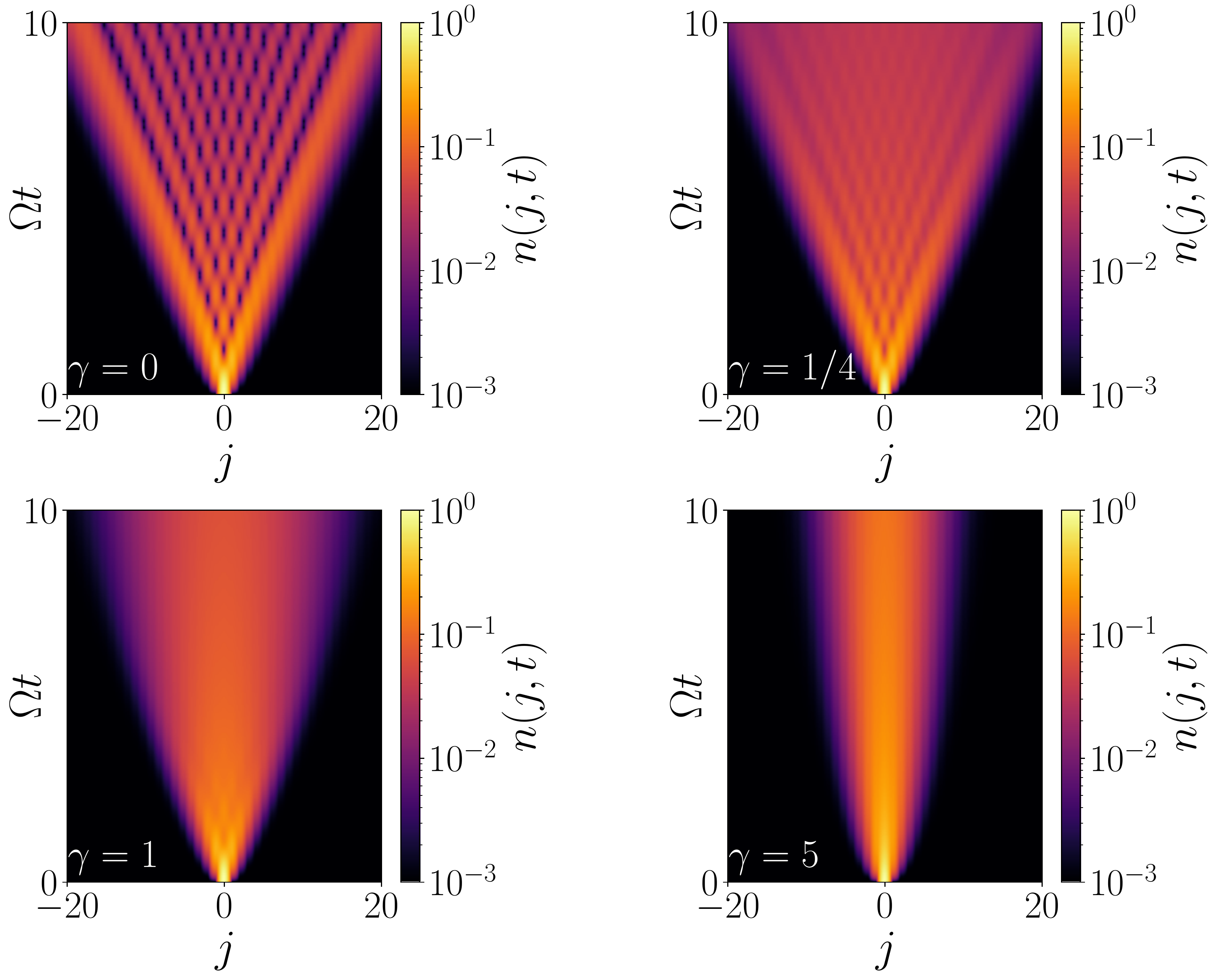}\\    \vspace{1cm}
    \includegraphics[width=\linewidth]{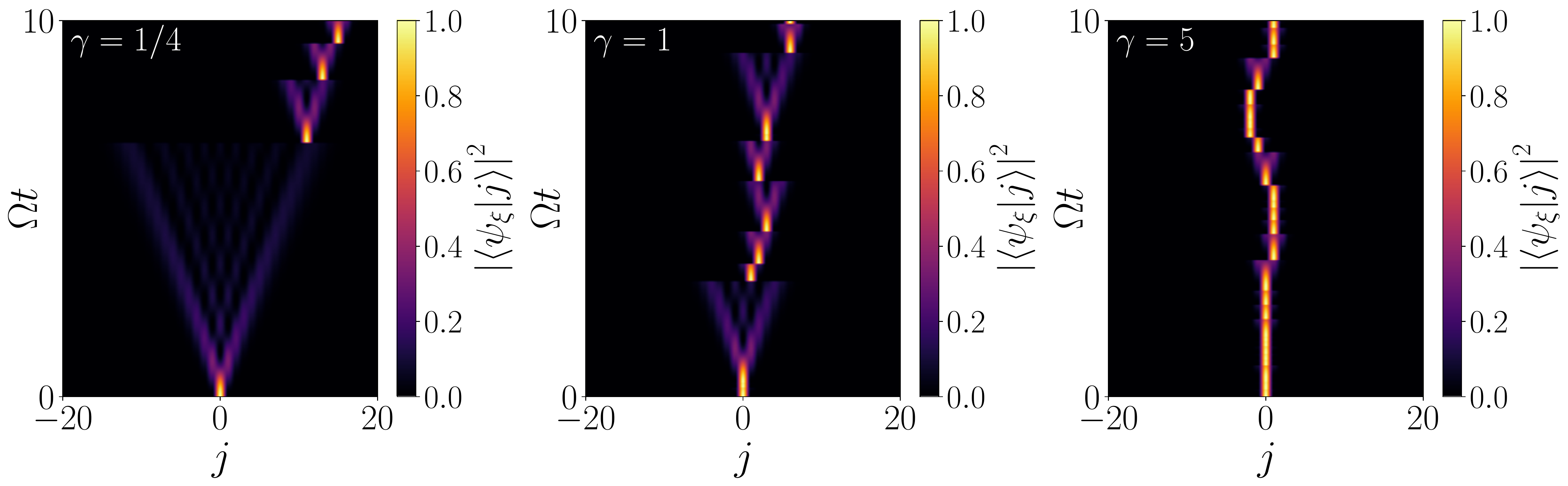}
    \caption{Upper four panels: average 
    local particle density $n(j,t)$ obtained from the average state given by Lindblad equation (Eq.~\ref{eq:lindblad2}). Different measurement rates are considered (i.e.\ $\gamma=0,0.25,1,5$), showing a clear transition from a ballistic to a diffusive dynamics. Lower three panels: the local particle density $|\langle \psi_{\xi}|j \rangle|^2$ obtained from quantum trajectories at different measurement rates ($\gamma=0.25,1,5$). }
    \label{fig:LindBlad_densities}
\end{figure}

In the case of the hopping particle, to analyze the dynamical map averaged over quantum trajectories, we can again reframe the measurement procedure as a Lindblad equation for the averaged density matrix $\rho(t)$. When we perform projective measurements of the particle's position at a rate $\gamma$, the average state $\rho$ undergoes the following transformation in accordance with the usual rules of quantum mechanics
\begin{equation}
\rho(t) \to \left(1-\gamma \dd t\right)\rho(t)+\gamma \dd t\sum_j \pi_j \rho(t) \pi_j \, .
\end{equation}
Here, $\pi_j = \dyad{j}$ represents the projectors over the lattice sites. The Lindblad master equation can be obtained by taking the limit $\dd t \to 0$ with a fixed value of $\gamma$, resulting in
\begin{equation}\label{eq:lindblad2}
    \partial_t \rho = -i \comm{H}{\rho}+\gamma\left[\sum_j \pi_j \rho \pi_j - \rho\right],
\end{equation}
with $H$ as in Eq.~\ref{eq:hoppins_ham}. In the case of a system with finite size, we can solve numerically the dynamical map\begin{equation}
    \rho(t+\dd t) = \rho(t)-i\dd t \left(H\rho-\rho H\right)+\gamma \dd t \left[\sum_j \pi_j \rho(t) \pi_j -\rho(t)\right],
\end{equation}
and define the site densities $n(j,t)=\Tr{\rho(t)\pi_j}$ as depicted in the upper four panels in Fig.~\ref{fig:LindBlad_densities}. In accordance to what have been discussed in the main text, the averages over the probability distribution coincides with the expectation values taken over the averaged state.

Now, we collect few results for the first moment of some relevant observables. In particular, from the probability distribution of a generic power $q^m$, by using the generic equation~\eqref{eq:moment1} we explicitly get
\be
\moment{}{q^m} = \Tr{\rho(t)q^m} =
e^{-\gamma t} 
\int_{-\pi}^{\pi} \frac{\dd k}{2\pi}
\sum_{j} j^m e^{-i j k}
I_{k}(t) \, .
\ee
By exploiting $\sum_{j} j^m e^{-i j k} = i^m 2\pi \delta^{(m)}(k)$,
we obtain
\be
\moment{}{q^m} = i^m e^{-\gamma t} \partial^{m}_{k} I_k(t)|_{k=0}.
\ee
As expected, $\moment{}{q} = 0$, while the first non-vanishing average occurs for $m=2$, giving
\be
\moment{}{q^2} = 
\frac{4 \Omega^2 \left(\gamma  t+e^{-\gamma t}-1\right)}{\gamma ^2}
\sim \begin{dcases}
        2 \Omega^2 t^2 \quad \gamma t \ll 1\\
        \frac{4\Omega^2t}{\gamma}
 \quad \gamma t \gg 1
    \end{dcases}
\ee
thus, we recover the ballistic behavior at early time (or for $\gamma = 0$), whilst a diffusive behavior for any $\gamma\neq 0$ at large time, with a diffusion constant $D_\gamma = 2\Omega^2/\gamma$ such that $\mean{x_{q_2}} \sim 2D_{\gamma}t$. This behavior has been observed in the many-body description of the model, in particular studying the particle current after quenching an initial domain wall configuration~\cite{DeLuca2019}. Due to its linear nature, this exact result can be compared with the solution of the Lindblad equation. Indeed, the average of $q^2$ is also given by
$\moment{}{q^2} = \sum_j j^2 n(j,t)$, where $n(j,t)=\Tr{\rho(t)\pi_j}$ can be evaluated numerically (see Fig.~\ref{fig:LindBlad_densities} upper panels), or analytically
taking the average over the probability distribution associated to $\pi_j$
\begin{equation}
    P_{\pi_j}(x;t) = \overline{\delta\left(\expval{\pi_j}{\psi_\xi(t)}-x\right)}.
\end{equation}
Explicitly we get
\be
\moment{}{\pi_j}
  = n(j,t) = e^{-\gamma t }
  \int_{-\pi}^\pi \frac{\dd k}{2\pi} e^{-ijk}I_k(t),
\ee
notice that, as expected, $\sum_j n(j,t)=1$, since 
$I_0(t) = e^{\gamma t}$.

\printbibliography
\end{document}